\documentclass[twocolumn,showpacs,preprintnumbers,amsmath,amssymb,latexsym,prd,footinbib,floatfix]{revtex4-2}

\usepackage{amssymb,amsmath}
\usepackage{graphicx}
\usepackage{multirow}  

\newcommand{\f}[2]{\frac{#1}{#2}}
\newcommand{\tf}[2]{{\textstyle\f{#1}{#2}}}
\newcommand{\la}{\langle}
\newcommand{\ra}{\rangle}
\newcommand{\de}{\partial}
\renewcommand{\Re}{{\rm Re}\,}
\renewcommand{\Im}{{\rm Im}\,}
\newcommand{\tr}{{\rm tr}\,}
\newcommand{\Oc}{{\mathcal O}}

\begin{document}

\title{Deconfinement transition and localization of Dirac modes in
  finite-temperature $\mathbb{Z}_3$ gauge theory on the lattice}

\author{Gy{\"o}rgy Baranka}
\email{barankagy@caesar.elte.hu}
\affiliation{ELTE E\"otv\"os Lor\'and University, Institute for
  Theoretical Physics, P\'azm\'any P\'eter s\'et\'any 1/A, H-1117, Budapest,
  Hungary}

\author{Matteo Giordano}
\email{giordano@bodri.elte.hu}
\affiliation{ELTE E\"otv\"os Lor\'and University, Institute for
  Theoretical Physics, P\'azm\'any P\'eter s\'et\'any 1/A, H-1117, Budapest,
  Hungary}

\begin{abstract}
  We study the localization properties of the eigenmodes of the
  staggered Dirac operator across the deconfinement transition in
  finite-temperature $\mathbb{Z}_3$ pure gauge theory on the lattice
  in 2+1 dimensions. This allows for nontrivial tests of the
  sea-islands picture of localization, according to which low modes
  should localize on favorable Polyakov-loop fluctuations in the
  deconfined phase of a gauge theory. We observe localized low modes
  in the deconfined phase of the theory, both in the real
  Polyakov-loop sector, where they are expected, and in the complex
  Polyakov-loop sectors, where they are not. Our findings expose the
  limitations of the standard sea-islands picture, and call for its
  refinement. An improved picture, where spatial hopping terms play a
  more prominent role, is proposed and found to be in excellent
  agreement with numerical results.
\end{abstract}

\maketitle

\section{Introduction}
\label{sec:intro}

Confinement of static color charges is one of the most striking
features of pure gauge theories, present at zero and low temperatures
for a large variety of gauge groups. While an analytic understanding
is still largely incomplete, this phenomenon has been convincingly
demonstrated by means of numerical simulations in lattice gauge
theory. However, the general mechanism of confinement, and of the
deconfinement transition observed at finite temperature, is still the
object of active research. A relatively recent approach to this issue
is through the study of the localization properties of the eigenmodes
of the Dirac operator, which are closely related to the confining
properties of the theory (see Ref.~\cite{Giordano:2021qav} for a
recent review).  In all the pure gauge theories examined so far, all
displaying an exact center symmetry, it was found that localized modes
are absent in the low temperature, confined phase, and present in the
high temperature, deconfined phase when the trivial Polyakov-loop
sector is selected, appearing exactly at the deconfinement transition
(within numerical
errors)~\cite{Gockeler:2001hr,Gattringer:2001ia,Gavai:2008xe,
  Kovacs:2009zj,Kovacs:2010wx,Bruckmann:2011cc,Kovacs:2017uiz,
  Giordano:2019pvc,Vig:2020pgq,Bonati:2020lal,Baranka:2021san}. The
connection between localization and deconfinement has been
demonstrated also in the presence of fermions, when a sharp transition
is present~\cite{Giordano:2016nuu,Cardinali:2021fpu}. Most
interestingly, this connection has been demonstrated, albeit in a
weaker sense, also in real-world QCD where the transition is only a
crossover, with localized modes appearing in the temperature range
where both confining and chiral properties of the theory change
rapidly~\cite{GarciaGarcia:2006gr,Kovacs:2012zq,Dick:2015twa,
  Cossu:2016scb,Holicki:2018sms}.  Here localization of the low modes
could be the link that ties these properties together, providing a
mechanism that explains the improvement of the chiral symmetry
properties generally observed at deconfinement in gauge theories with
fermions (e.g., through the reduction of the chiral condensate).

A qualitative understanding of the close relationship between
localization and deconfinement has been suggested in
Ref.~\cite{Bruckmann:2011cc} and further developed in
Refs.~\cite{Giordano:2015vla,Giordano:2016cjs,Giordano:2016vhx,
  Giordano:2021qav}, and is referred to as the ``sea-islands picture''
of localization. In this picture, the localization of Dirac eigenmodes is explained in
terms of two features: (1) the presence in the high-temperature phase
of a "sea" of ordered local Polyakov loops that get close to 1 in the
physical, real center sector selected by fermions and (2) the presence
of "islands" of Polyakov-loop fluctuations away from the ordered
value. The main
effect of Polyakov-loop ordering, combined with the twist imposed on
fermion wave functions by the antiperiodic temporal boundary
condition, is to open a ``pseudogap'' in the spectral density of the
Dirac operator, driven by the lowest Matsubara frequency.  The effect
of a nontrivial Polyakov-loop fluctuation is to effectively and
locally reduce the temporal twist on the fermion wave function, and
for modes localized on the fluctuation this is expected to lower the
eigenvalue below the lowest Matsubara frequency, if the spatial
hopping terms do not offset the gain.  As long as this is the case, it
is then ``energetically'' convenient for the eigenmodes to localize on
islands of fluctuations, leading to populating the pseudogap with a
relatively low density of modes.

The sea-islands picture leads one to expect localized low modes in the
deconfined phase of a generic gauge theory where Polyakov loops get
ordered near the trivial value, independently of the gauge group and
of the dimensionality of the system. Such an expectation is supported
by numerical results covering a wide variety of gauge
theories~\cite{Giordano:2021qav,Gockeler:2001hr,
  Gattringer:2001ia,Gavai:2008xe,Kovacs:2009zj,Kovacs:2010wx,
  Bruckmann:2011cc,Kovacs:2017uiz,Giordano:2019pvc,Vig:2020pgq,
  Bonati:2020lal,Baranka:2021san,Giordano:2016nuu,GarciaGarcia:2006gr,
  Kovacs:2012zq,Dick:2015twa,Cossu:2016scb,Holicki:2018sms} and
related models~\cite{Giordano:2015vla,Giordano:2016cjs,
  Giordano:2016vhx,Bruckmann:2017ywh}. The observed correlation
between localized modes and Polyakov-loop fluctuations supports the
mechanism outlined above~\cite{Bruckmann:2011cc,Cossu:2016scb,
  Holicki:2018sms,Baranka:2021san}. These studies include also the
simplest theory displaying a deconfinement transition, i.e.,
$\mathbb{Z}_2$ gauge theory in 2+1 dimensions, investigated by us in
Ref.~\cite{Baranka:2021san}. The study of the simplest gauge models
can provide valuable information on the mechanisms underlying
localization, since in these models many features that should be
irrelevant to localization but could confuse the picture are simply
absent.

One should mention at this point that a second localization mechanism
is available for topologically nontrivial gauge groups. In this case
one expects to find near-zero modes of topological nature, originating
from the exact zero modes supported by isolated calorons and
anti-calorons. Since at high temperature (anti)calorons form a dilute
medium, the corresponding zero-modes can mix little with each other
(as well as with delocalized modes that are well separated in energy,
living beyond the pseudogap), and so topological near-zero modes are
expected to be localized (see Refs.~\cite{Diakonov:1995ea,
  GarciaGarcia:2005vj}). A peak of localized near-zero modes has been
indeed observed in the spectral density in SU(3) gauge theory in 3+1
dimensions~\cite{Vig:2021oyt}. (See, however, also
Refs.~\cite{Alexandru:2021pap,Alexandru:2021xoi} for a different point
of view on the behavior of the lowest, almost-zero modes.)  A similar
peak is present also in QCD with near-physical and lower-than-physical
quark masses~\cite{Alexandru:2015fxa,Dick:2015twa,Ding:2020xlj,
  Kaczmarek:2021ser}, and there are indications that these modes are
localized~\cite{Dick:2015twa}, which could have interesting
consequences if the peak survives the chiral
limit~\cite{giordano_GT_lett,Giordano:2021nat,Giordano:2022ghy}. In
spite of appearance, this mechanism is not in contrast with the
sea-islands picture, since the Polyakov loop is non-trivial near
(anti)calorons, but complements it by indicating a source of favorable
Polyakov-loop fluctuations when there is non-trivial topology. On the
other hand, localization has been observed also when topology is
trivial; even when it is non-trivial, topological fluctuations are not
sufficient to account for all the localized
modes~\cite{Bruckmann:2011cc,Kovacs:2019txb}. The sea-islands
mechanism then appears to be more fundamental.

The sea-islands picture can be extended to the case where a nontrivial
Polyakov-loop sector is selected in the deconfined phase, provided one
takes into account that here the ordered Polyakov loop does not
correspond to the maximal possible twist for the fermions.  This leads
to a variety of scenarios. For example, if Polyakov loops get ordered
near $-1$, as may be the case, e.g., in ${\rm SU}(2N)$ or
$\mathbb{Z}_{2N}$ theories, then localization of low modes is not
expected, since the ordered loops already correspond to the most
favorable places, where the twist on the fermion wave function is
completely offset by the Polyakov loop, and all fluctuations
correspond to a non-vanishing twist. This has been verified in 2+1
dimensional $\mathbb{Z}_2$ gauge theory~\cite{Baranka:2021san}. For
QCD at imaginary chemical potential $\mu_I = \pi T$, one of the
complex Polyakov-loop sectors $e^{\pm i\f{2\pi}{3}}$ is selected above
the deconfinement (Roberge-Weiss) temperature, and local fluctuations
to the real sector reduce the twist on the fermion wave function. This
leads one to expect localization of the low modes, that has been
observed in Ref.~\cite{Cardinali:2021fpu}.

In this context, a particularly interesting setup is the deconfined
phase of $\mathbb{Z}_{N}$ theories with $N$ odd, when Polyakov loops
get ordered near $-e^{\mp i\f{\pi}{N}}$. In this case one finds again
that the sea of ordered loops corresponds to the most favorable twist
on the fermion wave function. At best, islands of fluctuations where
the Polyakov loop takes the value $-e^{\pm i\f{\pi}{N}}$ can provide
an equally but not more convenient twist, and so one is led to expect
delocalized low modes. Finding localized modes instead would pose a
challenge to the standard sea-islands picture of localization, and
would require the effect of spatial hopping terms in the staggered
operator to be favorable to localization, contrary to what one would
naively expect.

In this paper we continue our study of localization of the low Dirac
modes and of the sea-islands picture investigating the second simplest
gauge theory with a deconfining phase transition, namely lattice
$\mathbb{Z}_3$ gauge theory in 2+1 dimensions, that we probe with the
staggered Dirac operator. There is a number of features that make this
model interesting. The most evident one is that since $\mathbb{Z}_3$
is the center of SU$(3)$, which is the gauge group of QCD, any insight
obtained here could be useful to better understand the physically
relevant case. A less evident feature is that $\mathbb{Z}_3$ is the
Abelian group where the standard sea-islands picture in the trivial
Polyakov-loop sector has the largest chance to fail. In fact, the
maximal possible gain in temporal twist provided by a Polyakov loop
fluctuating to one of the complex, non-trivial sectors in a sea of
trivial Polyakov loops is here at its lowest, and the effect of the
spatial hopping terms might prevent localization. The least evident
and most interesting feature is that, as pointed out above, when the
Polyakov loop gets ordered in one of the complex sectors one has no
reason to expect localization of the low modes based on ``energetic''
considerations. This provides a nontrivial test of the standard
sea-islands picture: indeed, finding localized low modes in this case
would require one to reconsider or at least refine it.

Quite surprisingly, while there is little doubt about its existence,
the deconfinement transition in $\mathbb{Z}_3$ gauge theory in 2+1
dimensions has not been previously studied in detail. While there are
extensive studies in the literature concerning
$\mathbb{Z}_2$~\cite{Caselle:1995wn} and
$\mathbb{Z}_{N\ge 5}$~\cite{Borisenko:2011aa,Borisenko:2012na,
  Borisenko:2012nf,Borisenko:2014vva}, no determination of the
critical temperature has been done for $\mathbb{Z}_3$. As a
preliminary task we then need to determine the deconfinement
temperature. This is most efficiently done exploiting the duality with
the 3-color Potts model (see, e.g, Ref.~\cite{Wipf:2013vp}), which
allows one to employ a straightforward cluster
algorithm~\cite{Swendsen:1987ce,Wolff:1988uh} for the numerical
simulations. (Incidentally, the critical temperature for the
$\mathbb{Z}_4$ model is straightforwardly obtained by exploiting its
equivalence with the $\mathbb{Z}_2\times \mathbb{Z}_2$ model at half
of the coupling, and is simply twice that of the $\mathbb{Z}_2$ model,
determined in Ref.~\cite{Caselle:1995wn}. The equivalence of the
$\mathbb{Z}_4$ and $\mathbb{Z}_2\times \mathbb{Z}_2$ models follows
from their being dual respectively to the four-state clock model and
to a decoupled pair of Ising models~\cite{Wipf:2013vp}, and from the
equivalence of these two models~\cite{suzuki1967}.)

The plan of the paper is the following. In Section \ref{sec:z3intro}
we briefly review $\mathbb{Z}_3$ gauge theory in 2+1 dimensions,
focussing in particular on its duality with the 3-color Potts model,
while in Section \ref{sec:loc} we briefly review localization of Dirac
modes and how it can be detected. In Section \ref{sec:DAham} we
revisit the standard sea-islands picture and its formulation in the
language of the ``Dirac-Anderson Hamiltonian'' for staggered fermions,
and provide a refined picture that better appreciates the role of the
spatial hopping terms. In Section \ref{sec:deconf} we report our
results on the deconfinement transition of $\mathbb{Z}_3$, concerning
in particular the critical temperature and the nature of the
transition. In Section \ref{sec:numloc} we study the localization
properties of staggered Dirac modes, in both phases of this model and,
in the deconfined phase, both in the real and in the complex
Polyakov-loop sectors, testing in particular the expectations of the
standard and of the refined sea-islands pictures. Finally, in Section
\ref{sec:concl} we draw our conclusions and show prospects for the
future. A few technical details related to duality and to the
sea-islands picture are discussed in Appendices \ref{sec:app_dual} and
\ref{sec:seaislands}, respectively.

\section{$\mathbb{Z}_3$ lattice gauge theory in 2+1 dimensions}
\label{sec:z3intro}

The Wilson action $S_{\mathbb{Z}_3}$ for finite-temperature
$\mathbb{Z}_3$ lattice gauge theory in 2+1 dimensions and the
corresponding partition function $Z_{\mathbb{Z}_3}$ read
\begin{equation}
  \label{eq:z3_1}
  \begin{aligned}
    Z_{\mathbb{Z}_3}(\beta) &= \sum_{\{U_\mu(n)\}} e^{-S_{\mathbb{Z}_3}[U;\beta]}\,,\\
    S_{\mathbb{Z}_3}[U;\beta] &=
    \beta\sum_{n}\sum_{\substack{\mu,\nu=1\\ \mu<\nu}}^3 (1-\Re
    U_{\mu\nu}(n))\,,
  \end{aligned}
\end{equation}
where $n=(n_1,n_2,n_3)=(\vec{x},t)$ runs over the sites of a cubic
${\cal V}=N_1N_2N_3=VN_3$ lattice, $n_\mu=0,\ldots,N_\mu-1$;
$\hat{\mu}$ denotes the unit lattice vector in direction $\mu$; the
sum is over all configurations of link variables $U_\mu(n)$ taking
values in $\mathbb{Z}_3$, $U_\mu(n)=e^{i\f{2\pi}{3}k_\mu(n)}$, with
$k_\mu(n)=0,1,2$; and $U_{\mu\nu}(n)$ are the plaquette variables
associated with elementary squares of the lattice,
\begin{equation}
  \label{eq:plaquette_def}
  U_{\mu\nu}(n)=U_\mu(n)U_\nu(n+\hat{\mu})U_{-\mu}(n+\hat{\mu}+\hat{\nu})U_{-\nu}(n+\hat{\nu})\,,
\end{equation}
where $U_{-\mu}(n) = U_\mu(n-\hat{\mu})^*$. Periodic boundary
conditions are imposed in all directions. At finite temperature, the
``temporal'' extension $N_t=N_3$ is kept fixed while the ``spatial''
extensions $N_{1,2}$ are eventually sent to infinity, typically
setting $N_s=N_1=N_2$. In terms of the (mass-dimension $1/2$) gauge
coupling $e$ and of the lattice spacing $a$, one has $\beta=1/(e^2a)$,
and so the temperature of the system is $T/e^2 = \beta/N_t$.

\subsection{Duality}
\label{sec:z3intro_duality}

For a lattice of infinite size, the partition function of the
$\mathbb{Z}_3$ gauge theory can be recast as that of a 3-state clock
(or vector Potts) model (see, e.g, Ref.~\cite{Wipf:2013vp}). This is
true also for a lattice of finite size, provided one sums over all
choices of cyclically shifted boundary conditions, i.e.,
\begin{equation}
  \label{eq:z3_3}
  Z_{\mathbb{Z}_3}(\beta) =
  z(\tilde{\beta})
  \sum_{\{B_\mu\}}Z_{\rm clock}^{\{B_\mu\}}(\tilde{\beta})\,,
\end{equation}
where $z(\tilde{\beta})$ is a numerical prefactor, while the partition
functions $Z_{\rm clock}^{\{B_\mu\}}$,
\begin{equation}
  \label{eq:z3_4}
  \begin{aligned}
    Z^{\{B_\mu\}}_{\rm clock}(\tilde{\beta}) &= \sum_{\{s(n)\}}
    e^{-S^{\{B_\mu\}}_{\rm clock}[s;\tilde{\beta}]}\,,\\
    S^{\{B_\mu\}}_{\rm
      clock}[s;\tilde{\beta}]&=\tilde{\beta}\sum_n\sum_{\mu=1}^3
    \left(1-\Re s(n)s(n+\hat{\mu})^*\right)\,,
  \end{aligned}
\end{equation}
describe the interaction of complex spin variables
$s(n)=e^{i\f{2\pi}{3}\sigma(n)}$, $\sigma(n)=0,1,2$, with boundary
conditions $\{B_\mu\}$,
\begin{equation}
  \label{eq:z3_6}
  s(n+N_\mu\hat{\mu})=B_\mu s(n)\,,\qquad \mu=1,2,3\,,
\end{equation}
where $B_\mu=e^{i\f{2\pi}{3}b_\mu}$, $b_\mu=0,1,2$. In
Eq.~\eqref{eq:z3_3} the dual coupling $\tilde{\beta}$ is set to
\begin{equation}
  \label{eq:z3_5}
  e^{\f{3\tilde{\beta}}{2}} = \f{1+2e^{-\f{3\beta}{2}}}{1-e^{-\f{3\beta}{2}}}\,.
\end{equation}
The need to sum over suitable boundary conditions to have an exact
duality in a finite volume is well known~\cite{gruber1977group,
  Caselle:1995wn, Caselle:2001im,vonSmekal:2012vx}. A simple general
argument showing that shifted boundary conditions are needed for
$\mathbb{Z}_N$ gauge theory is given in Appendix \ref{sec:app_dual}.
The effect of such boundary conditions has been discussed in
Refs.~\cite{Caselle:1995wn,Caselle:2001im,vonSmekal:2012vx}.  The
presence of nontrivial $B_{1,2}$ in the spatial boundary conditions
leads only to finite-size corrections to the free energy with respect
to the trivial case. For the temporal boundary conditions, a
nontrivial $B_3$ leads in the ordered phase to the formation of a
spacelike interface between differently ordered domains, and so to an
increase in the corresponding free energy and a suppression of the
corresponding partition function $Z_{\rm clock}^{\{B_\mu\}}$. In the
disordered phase, instead, a nontrivial $B_3$ leads only to
finite-size corrections, and so all $Z_{\rm clock}^{\{B_\mu\}}$ are
equal in the thermodynamic limit.  In both phases one can then
restrict to $Z_{\rm clock}\equiv Z^{\{B_\mu=1\}}_{\rm clock}$ and
obtain the correct $V\to\infty$ limit for thermodynamic observables.
It is worth mentioning that the 3-state clock model is equivalent to
the 3-color Potts model,
\begin{equation}
  \label{eq:z3_7}
  \begin{aligned}
    Z_{\rm clock}(\tilde{\beta}) &= Z_{\rm
      Potts}(\tf{3}{2}\tilde{\beta})\,,\\
    Z_{\rm Potts}(\bar{\beta}) &=\sum_{\{s(n)\}} e^{-S_{\rm
        Potts}[s;\bar{\beta}]}
    \,,    \\
    S_{\rm Potts}[s;\bar{\beta}] &=
    \bar{\beta}\sum_n\sum_{\mu=1}^3(1-\delta_{\sigma(n),\sigma(n+\hat{\mu})})
    \,.
  \end{aligned}
\end{equation}
This is actually true irrespectively of the dimension and of the
(matching) choice of boundary conditions.

\subsection{Critical behavior}
\label{sec:z3gauge_crit}

The 2+1 dimensional $\mathbb{Z}_3$ gauge theory is expected to display
a deconfinement transition at some critical $\beta_c=\beta_c(N_t)$,
where the local Polyakov loops,
\begin{equation}
  \label{eq:z3_8}
  P(\vec{x})   \equiv \prod_{t=0}^{N_t-1} U_3(\vec{x},t)\,,
\end{equation}
align to one of the center elements $e^{i\f{2\pi z}{3}}$, $z=0,1,2$
(of course an Abelian group coincides with its center), and the center
symmetry of the model under the transformation
\begin{equation}
  \label{eq:z3_9}
 U_3(\vec{x},N_t-1)\to e^{i\f{2\pi z}{3}}U_3(\vec{x},N_t-1)\,,~~ \forall \vec{x}\,, 
\end{equation}
breaks down spontaneously. The duality relation discussed above
implies that the critical behavior of this model is the same as that
of the 3-color Potts model in a thin-film, 2+1 dimensional geometry,
which in turn is expected to match that of the corresponding
two-dimensional model. We then expect the deconfinement transition in
2+1 dimensional $\mathbb{Z}_3$ gauge theory to be second order, and in
the same universality class as that of the two-dimensional 3-color
Potts model, whose critical exponents are known (see, e.g.,
Ref.~\cite{Wu:1982ra}).

From the numerical point of view, it is convenient to determine
$\beta_c(N_t)$ by exploiting the duality relation and determining
instead the critical coupling $\bar{\beta}_c(N_t)$ of the 2+1
dimensional 3-color Potts model, for which one can use a cluster
algorithm~\cite{Swendsen:1987ce, Wolff:1988uh} and overcome the
critical slowing down of local update algorithms near the
transition. A convenient (complex) order parameter for the Potts model
is the quantity
\begin{equation}
  \label{eq:z3_10}
  \Phi = \f{1}{{\cal V}}\sum_n e^{i \frac{2 \pi }{3} \sigma(n)}\,,
\end{equation}
whose expectation value vanishes in the low-$\bar{\beta}$, disordered
phase and is nonzero in the high-$\bar{\beta}$, ordered phase. To
determine the critical coupling we performed a finite-size-scaling
study of the following Binder parameter~\cite{Binder:1981sa},
\begin{equation}
  \label{eq:z3_11}
  {\cal B} = \f{\la |\Phi|^4\ra }{\phantom{{}^2}\la |\Phi|^2\ra^2}\,,
\end{equation}
where $\la\ldots\ra$ denotes the expectation value associated with the
partition function $Z_{\rm Potts}$, Eq.~\eqref{eq:z3_7}. In the
disordered phase, in the large-volume limit $\Phi$ is expected to obey
a (two-dimensional) Gaussian distribution centered at the origin, and
so ${\cal B}\to 2$ as the system size $L=N_s$ tends to infinity. In
the ordered phase, instead, the distribution of $\Phi$ is peaked at a
non-zero value and ${\cal B}\to 1$ as $L\to \infty$.  Under the usual
one-parameter scaling hypothesis, near the critical coupling
$\bar{\beta}_c$ one has that ${\cal B}$ depends only on the ratio of
the (infinite-volume) correlation length
$\xi\sim |\bar{\beta}-\bar{\beta}_c|^{-\nu}$ and $L$,
${\cal B} = F\left(\xi/L\right)$. Since this must be an analytic
function of $\beta$ as long as $L$ is finite, one finds
\begin{equation}
  \label{eq:z3_12}
  {\cal B}(\bar{\beta},L) 
  = f\left((\bar{\beta}-\bar{\beta}_c)L^{\f{1}{\nu}}\right)\,,
\end{equation}
for some analytic function $f$, and so at $\bar{\beta}_c$ the Binder
parameter is scale invariant.

\section{Localization of Dirac eigenmodes}
\label{sec:loc}

In this Section we briefly discuss eigenmode localization and how to
detect it. Full accounts can be found in the literature (see, e.g.,
Refs.~\cite{lee1985disordered,Evers:2008zz,Giordano:2021qav}). In this paper we
investigate the localization properties of the eigenmodes of the
staggered Dirac operator,
\begin{equation}
  \label{eq:loc1}
  \begin{aligned}
    D^{\rm stag}_{n,n'} &= \f{1}{2}\sum_{\mu=1}^3
    \eta_\mu(n)\left(U_\mu(n)\delta_{n+\hat{\mu},n'}
      -U_{-\mu}(n)\delta_{n-\hat{\mu},n'}\right)\,, \\ \eta_\mu(n) &=
    (-1)^{\sum_{\nu<\mu}n_\nu}\,,
  \end{aligned}
\end{equation}
computed in the background of gauge field configurations obtained in
$\mathbb{Z}_3$ pure gauge theory.  Periodic boundary conditions in the
spatial directions and antiperiodic boundary conditions in the
temporal direction are understood. In this context $D^{\rm stag}$ acts
simply as a probe of the gauge dynamics, which does not include any
backreaction from the fermionic modes.

In the deconfined phase of the theory, the eigenmodes of
$D^{\rm stag}$ should be studied separately for the different center
sectors, characterized by the center element closest to the spatially
averaged Polyakov loop,
\begin{equation}
  \label{eq:pol_spav}
  \bar{P} \equiv \f{1}{V}\sum_{\vec{x}} P(\vec{x})\,. 
\end{equation}
One can in fact imagine including very heavy dynamical staggered
fermions, which explicitly break the center symmetry of the theory and
favor the trivial center sector, and then remove them by sending their
mass to infinity. In this limit center symmetry is not broken
explicitly, but it is broken spontaneously in the deconfined phase,
with the trivial center sector being selected by the procedure
outlined above. The same procedure but in the presence of a suitable
imaginary chemical potential selects instead one of the complex
sectors. In practice, in the deconfined phase one simply studies the
eigenmodes of $D^{\rm stag}$ restricting to configurations in the
center sector of interest.  The change in the properties of the
eigenmodes of $D^{\rm stag}$ as the pure gauge system transitions from
the confined phase to the deconfined phase in a specific center sector
then reflects how (infinitely) heavy staggered fermions see the phase
transition (see Ref.~\cite{Baranka:2021san} for a more detailed
discussion).

Since $D^{\rm stag}$ is anti-Hermitian, its eigenmodes have purely
imaginary eigenvalues, $D^{\rm stag}\psi_l = i\lambda_l\psi_l$,
$\lambda_l\in \mathbb{R}$. Moreover, the spectrum $\{\lambda_l\}$ is
symmetric about the origin due to the chiral property
$\{\eta_5,D^{\rm stag}\}=0$, where
$\eta_5(n) = (-1)^{\sum_{\nu}n_\nu}$, so that
$D^{\rm stag}\eta_5\psi_l = -i\lambda_l\eta_5\psi_l$. Since also the
eigenmode amplitude squared, $|\psi_l(n)|^2$, is the same for $\psi_l$
and $\eta_5\psi_l$, it suffices to restrict our attention to
$\lambda_l\ge 0$.  It is understood that eigenmodes are normalized,
$\sum_n|\psi_l(n)|^2=1$.

\subsection{Participation ratio}
\label{sec:loc_pr}

The staggered operator is technically ($-i$ times) the Hamiltonian of
a disordered system, with disorder provided by the fluctuations of the
gauge links. For gauge theories with a mass gap, disorder (i.e., gauge
field) correlations are short-ranged. Such systems are well known in
the condensed matter community to display eigenmode localization,
typically at the spectrum edge~\cite{lee1985disordered,Evers:2008zz}.
Whether eigenmodes in a given spectral region are localized or not can
be determined quantitatively by studying the volume scaling of their
participation ratio (PR),
\begin{equation}
  \label{eq:loc2}
  {\rm PR}_l\equiv \f{1}{N_t V}{\rm IPR}_l^{-1}\,,\quad  {\rm IPR}_l
  \equiv \sum_n|\psi_l(n)|^4\,, 
\end{equation}
averaged over configurations and locally in the spectrum. For a
generic observable $\Oc_l$ associated with mode $l$, we denote this
type of average as
\begin{equation}
  \label{eq:loc3}
  \overline{\Oc}(\lambda,V) \equiv 
  \f{\la    \sum_l\delta(\lambda-\lambda_l) \Oc_l\ra}{{\cal V}\rho(\lambda)}
  \,,
\end{equation}
where the dependence on the spatial volume is made explicit,
$\la\ldots\ra$ denotes the expectation value associated with the
partition function $Z_{\mathbb{Z}_3}$, Eq.~\eqref{eq:z3_1}, and
$\rho(\lambda)$ is the spectral density,
\begin{equation}
  \label{eq:loc3bis}
  \rho(\lambda) \equiv \f{1}{{\cal V}}\la
  {\textstyle\sum_l}\delta(\lambda-\lambda_l) \ra\,. 
\end{equation}
The PR effectively measures the fraction of the system occupied by an
eigenmode.  As $V\to \infty$, in a spatially two-dimensional system
one expects
\begin{equation}
  \label{eq:loc3ter}
  \overline{\rm PR}(\lambda,V) \sim
  c(\lambda)V^{\f{\alpha(\lambda)-2}{2}}\,,
\end{equation}
for some $0\le \alpha(\lambda)\le 2$, which is referred to as fractal
dimension. If modes in a certain spectral region are localized, i.e.,
if they typically extend over a finite region whose size does not
scale with the lattice size, then $\alpha=0$ in that region. If
instead they keep spreading out as the system size increases, then
they are delocalized and $\alpha\neq 0$. In particular, for $\alpha=2$
the modes spread out at the same speed as the system size increases,
and so are fully delocalized throughout the system.  In the condensed
matter literature, ``delocalized'' is usually reserved for $\alpha=2$,
while modes with $0<\alpha<2$ are called ``critical''.  Instead of the
PR, one can equivalently look at the average ``size'' of the modes,
$N_t V \cdot \overline{\rm PR}(\lambda,V)=\overline{{\rm
    IPR}^{-1}}(\lambda,V)$. In the large-volume limit this quantity
tends to a constant for localized modes, while it diverges for
delocalized modes.

\subsection{Localization and spectral statistics}
\label{sec:loc_sstat}

Another way to detect localization is by looking at the statistical
properties of the spectrum, which reflect the localization properties
of the eigenmodes~\cite{altshuler1986repulsion}.  Fluctuations of
delocalized modes under changes in the gauge field configuration are
strongly correlated, and so the corresponding eigenvalues are expected
to obey the appropriate type of Random Matrix Theory (RMT)
statistics. On the other hand, localized modes are uncorrelated, as
they respond essentially only to variations of the gauge fields where
they are localized, and so the corresponding eigenvalues are expected
to fluctuate independently and obey Poisson statistics. This is most
easily revealed by the unfolded spectrum, defined via the following
mapping,
\begin{equation}
  \label{eq:loc_stat1}
  \lambda_i \rightarrow x_i = {\cal V} \int^{\lambda_i}d\lambda\,
  \rho(\lambda)\,,
\end{equation}
for which universal predictions are available both for Poisson and RMT
statistics (see, e.g., Ref.~\cite{mehta2004random}). In particular,
the probability distribution of the unfolded level spacings
$s_i=x_{i+1}-x_i$ is known analytically.  For Poisson statistics this
is the exponential distribution,
\begin{equation}
  \label{eq:loc_stat2}
  p_{\rm P}(s)=e^{-s}\,.
\end{equation}
For RMT statistics the distribution is known from the solution of the
Gaussian ensemble in the various symmetry classes, with the unitary
one being relevant in the case at hand, but no closed form is
available. A good approximation is provided by the so-called Wigner
surmise,
\begin{equation}
  \label{eq:loc_stat3}
  p_{\rm WS}(s)=a_\beta s^\beta e^{-b_\beta s^2},
\end{equation}
where $\beta=2$, $a_2=\frac{32}{\pi^2}$ and $b_2=\frac{4}{\pi}$ for
the unitary class. A transition from localized to delocalized modes in
the spectrum corresponds to a change in the local statistical
properties of the eigenvalues, which can be monitored by looking at
features of the unfolded level spacing distribution $p_\lambda(s)$
computed locally in the spectrum. A practically convenient choice is
the integrated probability distribution,
\begin{equation}
  \label{eq:loc_stat4}
  I_{s_0} (\lambda,V)=  
  \int_0^{s_0}ds\, p_\lambda(s) =
  \overline{\theta^{s_0}}(\lambda,V)\,,
\end{equation}
where $\overline{\theta^{s_0}}$ is the average in the sense of
Eq.~\eqref{eq:loc3} of the observable
$\theta^{s_0}_l= \theta(s_0-s_l)$, with $\theta(s)$ the Heaviside
function.  Here $s_0\approx 0.508$ is chosen as the first crossing
point of the exponential and the Wigner surmise to maximize the
difference between the Poisson and RMT expectations for this quantity,
which are respectively $I_{s_0}^{\rm P} \simeq 0.398$ and
$I_{s_0}^{\rm WS}\simeq 0.117$.

\subsection{Anderson transitions}
\label{sec:loc_AT}

Regions in the spectrum where modes are localized are separated from
regions where they are delocalized by ``mobility edges'', $\lambda_c$,
where the localization length diverges and the system undergoes a
phase transition along the spectrum. The nature of such transitions,
known as ``Anderson transitions'', depends on the dimensionality and
the symmetry class of the system~\cite{Evers:2008zz}. While in three
spatial dimensions the known Anderson transitions are all second order
(including in finite-temperature QCD~\cite{Giordano:2013taa}), in two
spatial dimensions one finds transitions of Berezinskii–Kosterlitz–Thouless (BKT) type for systems in the
orthogonal~\cite{zhang2009localization} and unitary
classes~\cite{xie1998kosterlitz,Giordano:2019pvc} (except at the
integer quantum Hall
transition~\cite{Zirnbauer:2018ooz,Dresselhaus:2021fbx}). There are
indications that an Anderson transition of BKT type is present also in
the spectrum of the staggered operator in lattice $\mathbb{Z}_2$ pure
gauge theory in 2+1 dimensions~\cite{Baranka:2021san}, which belongs
to the orthogonal class. In a BKT-type Anderson transition, the
localization length diverges exponentially in
$(\lambda-\lambda_c)^{-\f{1}{2}}$ at the mobility edge $\lambda_c$,
and modes change from localized to critical, i.e., delocalized but
with a nontrivial fractal dimension $0<\alpha<2$, that keeps changing
along the spectrum above $\lambda_c$.  This peculiar behavior affects
that of the spectral statistics: for BKT-type Anderson transitions the
spectral region beyond the mobility edge does not obey RMT statistics,
displaying instead a continuously varying statistics intermediate
between Poisson and RMT, reflecting the critical nature of the
eigenmodes (see Refs.~\cite{xie1998kosterlitz,Giordano:2019pvc} and
references therein). Since the staggered operator in the background of
$\mathbb{Z}_3$ gauge fields is in the unitary class, if an Anderson
transition is present one would then expect to observe the features of
a BKT-type transition.

\subsection{Correlations with gauge observables}
\label{sec:loc_gauge}

According to the sea-islands mechanism mentioned in the Introduction,
in the deconfined phase of a gauge theory one generally expects the
low Dirac modes to localize near local fluctuations of the Polyakov
loop away from the ordered value. This can be checked by measuring the
correlation $\overline{\cal P}(\lambda,V)$ between modes and Polyakov
loops, where
\begin{equation}
  \label{eq:loc4}
  {\cal P}_l
  \equiv     \sum_{\vec{x},t}
  P(\vec{x})|\psi_l(\vec{x},t)|^2\,,
\end{equation}
and the average is taken according to Eq.~\eqref{eq:loc3}.  For fully
delocalized modes one expects
$\overline{\cal P} \simeq \bar{P} = \f{1}{V}\sum_{\vec{x}}
P(\vec{x})$, while for localized modes, that should be concentrated on
islands of fluctuations, a clearly different value should be obtained.
Notice that in the confined phase, and in the deconfined phase in the
real sector, one expects $\Im\overline{\cal P} = 0$ due to
charge-conjugation invariance.

Another interesting correlation to check is that between modes and
nontrivial plaquettes. To this end we looked at the quantities
$\overline{\cal U}$ and $\overline{{\cal U}_*}$ obtained via
Eq.~\eqref{eq:loc3} from the following observables,
\begin{equation}
  \label{eq:loc5}
  \begin{aligned}
    {\cal U}_l &\equiv \sum_{n} A(n) |\psi_l(n)|^2 \,, &&& {\cal
      U}_{*l} &\equiv \sum_{\substack{n\\ A(n)> 0}} |\psi_l(n)|^2\,,
  \end{aligned}
\end{equation}
where 
\begin{equation}
  \label{eq:loc6}
  \begin{aligned}
    A(n) \equiv \f{2}{3}{\rm
      Re}{\sum_{\substack{\mu,\nu=1\\\mu<\nu}}^3} &[ 4 -
    U_{\mu\nu}(n)- U_{\mu\nu}(n-\hat{\mu}) \\
    & -U_{\mu\nu}(n-\hat{\nu}) - U_{\mu\nu}(n-\hat{\mu}-\hat{\nu})]
  \end{aligned}
\end{equation}
equals the number of nontrivial plaquettes touching $n$.  The quantity
$\overline{\cal U}$ then counts the average number of nontrivial
plaquettes seen by a mode, and so for delocalized modes one expects
$\overline{\cal U} \simeq 8 \la 1-U_{\mu\nu}\ra $. The quantity
$\overline{{\cal U}_*}$ instead measures how much of the mode weight
is found on the corners of nontrivial plaquettes. Both observables
measure, in slightly different ways, how sensitive eigenmodes are to
nontrivial plaquettes: $\overline{{\cal U}_*}$ shows in general how
much eigenmodes are attracted to or repelled from such plaquettes,
while $\overline{\cal U}$ shows how attracted eigenmodes are to
regions where nontrivial plaquettes cluster together.

\section{Sea-islands picture and the Dirac-Anderson Hamiltonian}
\label{sec:DAham}

A more detailed formulation of the sea-islands picture for staggered
fermions is based on the ``Dirac-Anderson Hamiltonian'' formalism,
developed in Ref.~\cite{Giordano:2016cjs} for non-Abelian theories.
While the adaptation to an Abelian theory is straightforward, we
review here the derivation in some detail since we are extending the
original analysis while adopting a slightly different point of view.
We work in $d+1$ dimensions for generality. A few technical details
are reported in Appendix \ref{sec:seaislands}.

\subsection{Dirac-Anderson Hamiltonian}
\label{sec:DAham1}

The Dirac-Anderson Hamiltonian $H^{\rm DA}$ is obtained via a unitary
transformation $\Omega$ as
$\Omega^\dag D^{\rm stag}\Omega = iH^{\rm DA}$, where $\Omega$ is the
matrix of spatially localized eigenvectors of the temporal part
($\mu=d+1$) of $D^{\rm stag}$. Eigenmodes are labelled by their
location $\vec{y}=(y_1,\ldots,y_d)$ and by an index
$k=0,\ldots,N_t-1$, corresponding to the $N_t$ Matsubara frequencies
$\omega_k(\vec{x})$ associated with the $\vec{x}$-dependent temporal
boundary condition
$\psi(\vec{x},N_t)=-P(\vec{x})\psi(\vec{x},0)=-e^{i\phi(\vec{x})}\psi(\vec{x},0)$. These
are discussed in detail below. For an Abelian theory with a
1-dimensional internal space one has
\begin{equation}
  \label{eq:DA_def1}
  \Omega_{t\vec{x},k\vec{y}} =
  \f{1}{\sqrt{N_t}}\delta_{\vec{x},\vec{y}}\,e^{i\omega_k(\vec{x})t}P(\vec{x},t)^*\,,
\end{equation}
where $P(\vec{x},t+1)= P(\vec{x},t) U_{d+1}(\vec{x},t)$, with
$P(\vec{x},0)=1$ and $ P(\vec{x},N_t) =P(\vec{x})=e^{i\phi(\vec{x})}$.
The corresponding Dirac-Anderson Hamiltonian reads
\begin{equation}
  \label{eq:DA_def2}
  \begin{aligned}
    H^{\rm DA}_{k\vec{x},l\vec{y}} &= e_k(\vec{x})\delta_{kl}
    \delta_{\vec{x},\vec{y}}    \\
    &\phantom{=}+\f{1}{2i}\sum_{j=1}^d\eta_j(\vec{x})
    [ V_{+j}(\vec{x})_{kl}(T_j)_{\vec{x},\vec{y}} \\[-1em]
    &\phantom{=+\f{1}{2i}\sum_{j=1}^d\eta_j(\vec{x})}-
    V_{-j}(\vec{x})_{kl}(T_j^\dag)_{\vec{x},\vec{y}}] \,,
  \end{aligned}
\end{equation}
with ``unperturbed eigenvalues''
\begin{equation}
  \label{eq:DA_def3}
  e_k(\vec{x})=    \eta_{d+1}(\vec{x})\sin\omega_k(\vec{x})\,,
\end{equation}
and hopping terms
\begin{equation}
  \label{eq:DA_def4}
  \begin{aligned}
    V_{\pm j}(\vec{x})_{kl}&=
    \f{1}{N_t}\sum_{t=0}^{N_t-1}e^{-i[\omega_k(\vec{x}) -
      \omega_l(\vec{x}\pm\hat{\jmath})]t} U_{\pm j}^{\rm
      tg}(\vec{x},t)\,,\\
    U_{\pm j}^{\rm tg}(\vec{x},t) &=P(\vec{x},t) U_{\pm j}(\vec{x},t)
    P(\vec{x}\pm \hat{\jmath},t)^*\,,
  \end{aligned}
\end{equation}
where $T_j$ are the spatial translation operators,
$(T_j)_{\vec{x},\vec{y}}=\delta_{\vec{x}+\hat\jmath ,\vec{y}}$
(including periodic boundary conditions). Here the superscript ``tg''
denotes links computed in temporal gauge,
$U_{d+1}^{\rm tg}(\vec{x},t)=1$, $\forall\vec{x}$, $0\le t< N_t-1$,
and we made explicit the fact that $\eta_{\mu}$ depend only on the
spatial coordinates. One has
$V_{-j}(\vec{x})_{kl}=V_{+j}(\vec{x}-\hat\jmath)_{lk}{}^*$, and one
can easily show that $V_{\pm j}(\vec{x})$ are unitary $N_t\times N_t$
matrices.

The Hamiltonian $H^{\rm DA}$ is identical to that of a set of $N_t$
Anderson-like models, with correlated random local potentials
$e_k(\vec{x})$, and coupled by the random hopping matrices
$V_j(\vec{x})$. At this stage, however, the labelling of the
$e_k(\vec{x})$ is arbitrary, and depends on the ordering in $k$ of the
basis vectors and on the convention chosen for the Polyakov-loop phase
$\phi(\vec{x})$, both of which can as well be
$\vec{x}$-dependent. This is formally expressed by writing
\begin{equation}
  \label{eq:DA_def5}
  \omega_k(\vec{x}) =
  \tilde{\omega}_{N_k(\vec{x})}(\phi(\vec{x}))\,,
  \quad
  \tilde{\omega}_n(\phi) = \f{\phi + (2n+1)\pi}{N_t}\,,
\end{equation}
where $N_k(\vec{x})=0,\ldots,N_t-1$ and the convention for the
Polyakov-loop phase have to be specified.  Without any loss of
generality, we can restrict the latter to
$\phi(\vec{x})\in [-\pi,\pi)$ at each site. Notice that since
$\sin \tilde{\omega}_{n+\f{N_t}{2}\;{\rm mod}\, N_t}= -
\sin\tilde{\omega}_n$, at each $\vec{x}$ half of the $e_k(\vec{x})$ in
Eq.~\eqref{eq:DA_def3} are positive and half are negative. A further
simplifying choice is then to pair opposite unperturbed eigenvalues so
that
\begin{equation}
  \label{eq:oppo}
   e_{k+\f{N_t}{2}{\rm mod}\,N_t}(\vec{x})=-e_{k}(\vec{x})\,.
\end{equation}
One can show that in this case
\begin{equation}
  \label{eq:oppo2}
  V_{\pm j}(\vec{x})_{k+\f{N_t}{2}{\rm mod}\,N_t\,\,l+\f{N_t}{2}{\rm
      mod}\, N_t}=V_{\pm j}(\vec{x})_{kl}\,.
\end{equation}
The resulting general structure of the Dirac-Anderson Hamiltonian is
then
\begin{equation}
  \label{eq:DA_general}
  \begin{aligned}
    H^{\rm DA} &=
    \begin{pmatrix}
      E & \mathbf{0} \\ \mathbf{0} & -E
    \end{pmatrix}
    \\ &\phantom{=} + \f{1}{2i}\sum_{j=1}^d \eta_j\left[
      \begin{pmatrix}
        A_j & B_j \\ B_j & A_j
      \end{pmatrix}T_j
      -T_j{}^\dag
      \begin{pmatrix}
        A_j{}^\dag & B_j{}^\dag \\ B_j{}^\dag & A_j{}^\dag
      \end{pmatrix}\right]\,,
  \end{aligned}
\end{equation}
where $E(\vec{x})$, $A_j(\vec{x})$, $B_j(\vec{x})$ are
$\f{N_t}{2}\times \f{N_t}{2}$ matrices, with $E$ diagonal,
$A_j{}^\dag A_j + B_j{}^\dag B_j = \mathbf{1}$ and
$ A_j{}^\dag B_j + B_j{}^\dag A_j =\mathbf{0}$.  Here $\mathbf{0}$ and
$\mathbf{1}$ are the $\f{N_t}{2}$-dimensional zero and identity
matrices, respectively.

The simplest possibility for $N_k(\vec{x})$ is clearly
$N_k^{(0)}(\vec{x})=k$, which with our convention on $\phi(\vec{x})$
satisfies the above requirements. This was used in
Ref.~\cite{Giordano:2016cjs}. However, the ordering of the $e_k(\vec{x})$ obtained with this choice
generally does not reflect their rank in magnitude. To give the $e_k(\vec{x})$ an intepretation
as the different ``energy levels'' of an electron in the potential of
an atom sitting at the spatial lattice site $\vec{x}$, it is
convenient to order the basis vectors so that $e_k(\vec{x})$ are
positive for $0\le k \le \f{N_t}{2}-1$ and negative for
$\f{N_t}{2}\le k \le N_t-1$, and in both cases ranked by absolute
value, i.e.,
\begin{equation}
  \label{eq:DA7_alt}
  0 \le e_0(\vec{x})\le e_1(\vec{x})\le \ldots \le
  e_{\f{N_t}{2}-1}(\vec{x})\,, 
\end{equation}
with $e_{\f{N_t}{2}+k}(\vec{x})=-e_{k}(\vec{x})$,
$k=0,\ldots,\f{N_t}{2}-1$. The diagonal matrix $E$ in
Eq.~\eqref{eq:DA_general} in this case is then positive-semidefinite.
This implicitly defines $N_k(\vec{x})$ so that a given $k$ always
corresponds to the energy level of the same rank at each spatial site,
thus giving $k$ an intrinsic meaning. An explicit expression for
$N_k(\vec{x})$ can be worked out analytically, and is reported in
Appendix \ref{sec:seaislands_rank}. In particular, the lowest positive
energy level at each site depends only on $\phi(\vec{x})$ and reads
\begin{equation}
  \label{eq:DA8_alt}
  e_0(\vec{x}) 
  =  {\cal E}(\phi(\vec{x}))=  \sin \f{\pi -
    |\phi(\vec{x})|}{N_t}  \,.
\end{equation}
Moreover, $e_k(\vec{x})$ remains unchanged under the replacement
$\phi(\vec{x})\to -\phi(\vec{x})$. 

Irrespectively of the choice of $N_k(\vec{x})$, the Dirac-Ander\-son
form of the staggered operator leads one to expect, by analogy with
the usual Anderson models, that modes at the high end of the spectrum
are localized, independently of the phase of the gauge system. Since
the ordering of the Polyakov loop at the deconfinement transition is
expected to open a pseudogap in the spectrum near the origin, thus
making the near-zero region qualitatively similar to a spectrum edge,
one can understand also localization of the low modes in the
deconfined phase by analogy with the usual Anderson models. However,
this intuitive explanation should be supplemented by a more detailed
mechanism if one wants to understand better the connection between
low-mode localization and deconfinement.

\subsection{Standard sea-islands picture}
\label{sec:SI_old}

As a first step in this direction, the formalism of the Dirac-Anderson
Hamiltonian discussed above allows one to formulate the sea-islands
picture more precisely. In fact, the twist on the fermion wave
functions provided by the effective local temporal boundary conditions
$\psi(\vec{x},N_t)=-P(\vec{x})\psi(\vec{x},0)$ can be quantified by
the lowest energy level ${\cal E}(\phi(\vec{x}))$,
Eq.~\eqref{eq:DA8_alt}. The sea-islands picture then amounts to state
that places with lower ${\cal E}$ are ``energetically'' favorable for
the localization of Dirac eigenmodes. This can be tested in detail by
looking at how much weight is allocated on the different ``branches''
of the wave function corresponding to the different energy levels, and
how this correlates with the energy levels themselves.

The components $\Psi_l(\vec{x},k)$ of the $l$th eigenvector in the new
basis are obtained from $\Psi_l=\Omega\psi_l$, and read
\begin{equation}
  \label{eq:app_DA6}
  \Psi_l(\vec{x},k) = \f{1}{\sqrt{N_t}}\sum_{t=0}^{N_t-1}
  e^{-i\omega_k(\vec{x}) t} P(\vec{x},t) \psi_l(\vec{x},t)\,.
\end{equation}
Clearly, $H^{\rm DA}\Psi_l = \lambda_l \Psi_l$. For each mode, the
weight on branch $k$ is
\begin{equation}
  \label{eq:DA9_alt}
  w_l(k) = \sum_{\vec{x}} |\Psi_l(\vec{x},k)|^2\,,
\end{equation}
and the corresponding unperturbed energy averaged over spatial
sites is
\begin{equation}
  \label{eq:DA10_alt}
  \varepsilon_l(k) = \f{1}{w_l(k)}\sum_{\vec{x}}e_k(\vec{x})
  |\Psi_l(\vec{x},k)|^2\,. 
\end{equation}
It is easy to show that the components $\Psi_{-l}(\vec{x},k)$ of
$\eta_5\psi_l$ (corresponding to eigenvalue $-\lambda_l$) in the new
basis are
\begin{equation}
  \label{eq:si_minuslam}
  \Psi_{-l}(\vec{x},k) =
  \eta_{d+1}(\vec{x})\Psi_{l}\left(\vec{x},k+\tf{N_t}{2} \,{\rm mod}\, N_t\right)\,,
\end{equation}
and so for the corresponding weights one finds
$w_{-l}(k)=w_l\left(k+\tf{N_t}{2} \,{\rm mod}\, N_t\right)$.

The sea-islands picture leads one to expect, perhaps na\"ively, that
low positive (resp.\ negative) modes, if present, have a large weight
$w_l(0)$ (resp.\ $w_l(\f{N_t}{2})$), and correspondingly
$\varepsilon_l(0)$ should be close to the most favorable value of
${\cal E}$ (resp.\ its negative), leading to localization when such
favorable places are rare.

\subsection{Refined sea-islands picture}
\label{sec:newsi}

While the idea of interpreting sites with low
${\cal E}(\phi(\vec{x}))$ as ``energetically'' fa\-vor\-able for the
eigenmodes is suggestive, the fact that $H^{\rm DA}$ is not
positive-definite makes it questionable. On the other hand, the
correlation between such sites and the localization centers of low
localized modes is evident in the numerical
data~\cite{Bruckmann:2011cc,Cossu:2016scb,Holicki:2018sms,
  Baranka:2021san}. We now argue that this correlation can be better
explained in an indirect way, which will lead us to a refinement of
the sea-islands mechanism.

In the high-temperature phase one expects the ordering of the Polyakov
loops to induce strong correlations among time slices. One then
expects to a first approximation that
$P(\vec{x})\approx P_*=e^{i\phi_*}$ and
$U_j^{\rm tg}(\vec{x},t)\approx U_{j*}^{\rm tg}(\vec{x})$, and so $E$
to be approximately $\vec{x}$-independent. An explicit calculation
[see Appendix \ref{sec:seaislands_soc}, Eq.~\eqref{eq:si3}] shows that
in this case
$V_{j}(\vec{x})_{kl} \approx \delta_{k,l+\f{N_t}{2}\,{\rm mod}\, N_t}
U_{j*}^{\rm tg}(\vec{x})$, or in other words
$A_j(\vec{x})\approx \mathbf{0}$ and
$B_j(\vec{x})\approx U_{j*}^{\rm tg}(\vec{x})\mathbf{1}$.  Notice that
this does not depend on the detailed definition of $N_k(\vec{x})$ as
long as Eq.~\eqref{eq:oppo} is enforced and the sign of $e_k(\vec{x})$
at fixed $k$ is constant throughout the lattice: The actual ordering
of the positive unperturbed eigenvalues is then immaterial, as it
should be. In this case
\begin{equation}
  \label{eq:onlyB}
  (H^{{\rm DA}})^2 \approx E^2 + \left[\f{1}{2i}\sum_{j=1}^d
    \eta_j\left(B_j T_j - T_j{}^\dag B_j{}^\dag \right)\right]^2
  \equiv H_B^2\,,
\end{equation}
and so generally a gap of size $[{\cal E}(\phi_*)]^2$ opens in the
spectrum of $(H^{{\rm DA}})^2$. In the opposite limit of
$B_j\approx \mathbf{0}$ and $A_j$ approximately proportional to
$\mathbf{1}$, again assuming constant $E$, one finds instead
\begin{equation}
  \label{eq:onlyA}
  \begin{aligned}
    (H^{{\rm DA}})^2 &\approx {\rm diag}\left(H_+^2,H_-^2\right)\equiv
    H_A^2\,,\\
H_\pm &= E \pm \f{1}{2i}\sum_{j=1}^d
\eta_j\left( A_j T_j - T_j{}^\dag A_j{}^\dag \right)\,,
  \end{aligned}
\end{equation}
which generally has an ungapped spectrum. One then generally expects
that low modes prefer locations where $A_j$ deviates from zero, and
even more so in the deconfined phase, where a large eigenmode
amplitude in these regions is required for the corresponding
eigenvalue to get below the gap. Since deviations from
$B_j^\dag B_j = \mathbf{1}$ are expected to show up in places where
the Polyakov loop is disordered, and since in the deconfined phase
these places are rare and spatially well separated from each other,
this makes them able to localize the low eigenmodes. While
$A_j\approx \mathbf{1}$ is unlikely to happen since it requires strong
\textit{anti}correlation among spatial links across different
time-slices [see Appendix \ref{sec:seaislands_soc},
Eq.~\eqref{eq:si3bis}], for Hamiltonians intermediate between
Eq.~\eqref{eq:onlyB} and Eq.~\eqref{eq:onlyA} -- $H_{AB}$-type
Hamiltonians, for want of a better name -- one still expects a
sizeable density of low modes.

To see this in detail, one can switch off the hopping terms connecting
the extended $B$-type region where $A_j\approx\mathbf{0}$ for all $j$,
from $AB$-type regions where it is non-negligible for some $j$, and
diagonalize separately the Hamiltonians of the resulting independent
subsystems. For the first region one finds a Hamiltonian of type
$H_B$, Eq.~\eqref{eq:onlyB}, and so delocalized modes and an almost
sharply gapped spectrum. For the second region one finds instead a
Hamiltonian of type $H_{AB}$ (or more precisely a set of spatially
separated Hamiltonians of this type), intermediate between
Eq.~\eqref{eq:onlyB} and Eq.~\eqref{eq:onlyA}, and so localized modes
and an ungapped spectrum, with a small but still sizeable density of
low modes. When switching on again the hopping terms at the boundary
of the two regions, one finds that the lowest $AB$-type modes, far
below the gap, can hardly mix with the $B$-type modes due to the large
energy difference, and with other $AB$-type modes due to the large
spatial separation, so that they remain localized also when accounting
for the full interaction.  This argument leads then to a different
type of sea-islands picture, where the sea and the islands are defined
in terms of the spatial hopping terms, rather than the local
potential.

The argument above clarifies the important role played by the ordering
of the Polyakov loop and the associated depletion of the near-zero
spectral region for the localization of the low modes in the
deconfined phase. By contrast, in the confined phase no extended
$B$-type region appears with its almost gapped spectrum, and there
seems to be no mechanism preventing delocalization of the low modes,
which are likely to display also a larger spectral density.

Notice also that for the same constant matrix $E$, the spectrum of the
Hamiltonian Eq.~\eqref{eq:onlyA} typically extends to higher values
than that of the Hamiltonian Eq.~\eqref{eq:onlyB}. An argument similar
to the one above suggests that places where $A_j$ deviates from zero
are favorable also for the localization of the high modes in the
deconfined phase, with delocalization being prevented by energy or
spatial separation.  In the confined phase where favorable islands are
more frequent, repeating the argument by separating now the regions
where $A_j$ is the largest from the rest, one sees that modes
localized on the most favorable islands are likely to reach larger
eigenvalues, and to remain stable against delocalization when the
remaining hopping terms are switched on due to a large energy
separation.

The refined sea-islands picture described above provides a more
detailed understanding of the microscopic mechanism behind the
localization of the low Dirac modes in the deconfined phase of a gauge
theory.  On the one hand, it does not contradict but rather subsume
the standard sea-islands picture in the physical, real Polyakov-loop
sector, since local disorder leading to $A_j\not\approx \mathbf{0}$ is
naturally associated with Polyakov-loop fluctuations, and these
automatically lead to a smaller ${\cal E}$. On the other hand, it
extends the old picture to cases where there really are no
``energetically favorable'' islands, such as the deconfined phase of
$\mathbb{Z}_3$ gauge theory in a complex center sector: even in this
case the local Polyakov-loop fluctuations are expected to lead to
favorable fluctuations in the hopping terms, despite the fact that
${\cal E}$ is not lower than in the sea of ordered Polyakov loops.
Moreover, fluctuations in the hopping terms could also appear
independently of the Polyakov-loop ones: according to the refined
picture these would be favorable for localization, while they are
overlooked by the standard picture.  This is a distinguishing
signature that can be looked for in numerical data.

As a final remark, we note that the argument above can be easily
extended to non-Abelian theories after dealing with only minor
technical complications: this is discussed in Appendix
\ref{sec:si_NA}.

\begin{figure}[t]
  \includegraphics[width=0.48\textwidth]{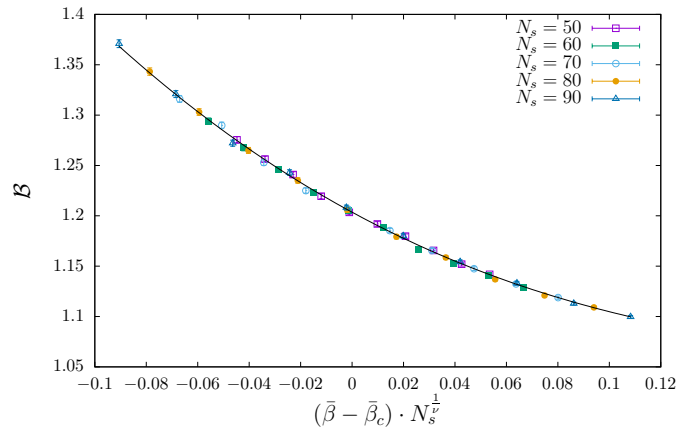}
  \caption{The Binder parameter ${\cal B}$, Eq.~\eqref{eq:z3_11}, for
    the 2+1 dimensional 3-color Potts model as a function of the
    coupling, for temporal size $N_t=4$ and various spatial sizes
    $N_s$.  The solid line shows the result of a fit to the data with
    Eq.~\eqref{eq:num_dec1}, for $n=6$.}
  \label{binder}
\end{figure}

\begin{table}[t]
  \begin{tabular}{c|c|c|c}
    \hline\hline
    $N_t$ & $N_s$ & configurations & fitting range \\ \hline\hline
    2 & 50,60,70,80,90 & 20000 & [0.6372,0.6381] \\\hline
    4 & 50,60,70,80,90 & 20000 & [0.5640,0.5649] \\\hline
    6 & 90,100,110,120,130 & 20000 & [0.5546,0.5557]
    \\\hline
  \end{tabular}
  \caption{Simulation and analysis details for the study of the 2+1
    dimensional 3-color Potts model for temporal extension $N_t$.}
  \label{tab:potts_stat}
\end{table}
\begin{table}[th]
  \begin{tabular}{c|cr@{.}lr@{.}lr@{.}lr@{.}l}
    \hline\hline    $N_t$ & & \multicolumn{2}{c}{$\bar{\beta}_c$} &
    \multicolumn{2}{c}{$\nu$} & \multicolumn{2}{c}{${\cal B}_c$} & \multicolumn{2}{c}{$\beta_c$}\\ \hline\hline
    \hline
    2 & & 0&637700(15)   & 0&772(63) & 1&1737(13) & 0&982070(16) \\ \hline
    4 & & 0&5644100(73)    & 0&846(30) & 1&2036(17) & 1&0670181(90) \\ \hline
    6 & & 0&555176(56) & 0&799(36) & 1&2246(28) & 1&078506(70)  \\ \hline\hline
  \end{tabular}
  \caption{Critical coupling $\bar{\beta}_c$, correlation length
    critical exponent $\nu$ and critical Binder parameter ${\cal B}_c$
    of the 2+1 dimensional 3-color Potts model for temporal extension
    $N_t$, and corresponding critical coupling $\beta_c$ of 2+1
    dimensional $\mathbb{Z}_3$ gauge theory.}
  \label{tab:potts}
\end{table}

\section{Deconfinement transition} 
\label{sec:deconf}

In this Section we report our numerical results on the deconfinement
transition in 2+1 dimensional $\mathbb{Z}_3$ gauge theory, determined
via duality from a study of the corresponding 3-color Potts model.
Periodic boundary conditions are understood on both sides of the
duality (see the discussion in Section \ref{sec:z3intro_duality}).

After a preliminary sweep with a standard Metropolis algorithm to
bracket the transition, we performed numerical simulations with a
standard cluster algorithm~\cite{Swendsen:1987ce,Wolff:1988uh} near
the critical coupling of the 2+1 dimensional 3-color Potts model on a
cubic $N_s^2\times N_t$ lattice. Keeping the temporal extension fixed
to $N_t=2,4,6$, we determined the critical coupling
$\bar{\beta}_c(N_t)$ by means of a finite-size-scaling analysis of the
Binder parameter ${\cal B}$, Eq.~\eqref{eq:z3_11}.  We measured
${\cal B}$ on a sample of well decorrelated configurations, estimating
its statistical error by a standard jackknife procedure. For the
finite-size-scaling analysis we made the usual one-parameter scaling
hypothesis, leading to Eq.~\eqref{eq:z3_12}, which we subsequently
approximated by a polynomial of order $n$,
\begin{equation}
  \label{eq:num_dec1}
  \begin{aligned}
    {\cal B}(\bar{\beta},N_s) &= \sum_{j=0}^n f_j\,
    u(\bar{\beta},N_s)^j
    \,, \\
    u(\bar{\beta},N_s)&=
    (\bar{\beta}-\bar{\beta}_c)N_s^{\f{1}{\nu}}\,.
  \end{aligned}
\end{equation}
We then fitted our numerical data with Eq.~\eqref{eq:num_dec1} using
the constrained fitting approach of Ref.~\cite{Lepage:2001ym}, varying
the order $n$ of the polynomial until the errors on the fitting
parameters stabilize. Through this procedure, the error estimate from
the fitting routine already takes into account the systematic effect
due to the truncation of Eq.~\eqref{eq:num_dec1}. Fits were performed
using the MINUIT library~\cite{James:1975dr,James:1994vla}. We did
not use any priors for $\bar{\beta}_c$ and $\nu$, and very broad
Gaussian priors for the coefficients $f_n$. Errors turn out to be
stable already at $n=6$. Details on the volumes, statistics, and
fitting ranges employed in the analysis are reported in
Tab.~\ref{tab:potts_stat}. Our results for the critical coupling
$\bar{\beta}_c$, the correlation length critical exponent $\nu$, and
the critical Binder parameter ${\cal B}_c = f_0$ are reported in
Tab.~\ref{tab:potts}. There we also report the dual critical coupling
$\beta_c$, at which the deconfinement transition takes place in
$\mathbb{Z}_3$ gauge theory. The quality of the resulting collapse
plot shown in Fig.~\ref{binder} confirms the goodness of the
one-parameter scaling assumption. The critical exponent $\nu$ is in
good agreement with the value $\nu^{\rm 2d\,Potts} = \f{5}{6}$ of the
2-dimensional 3-color Potts modes~\cite{Wu:1982ra}, confirming the
second-order nature and universality class of the transition expected
from universality arguments (see Section \ref{sec:z3gauge_crit}).  For
comparison, the critical parameters of the two-dimensional model are
$\beta_{c}^{\rm
  2d\,Potts}=\log(1+\sqrt{3})$~\cite{Wu:1982ra,Beffara2012} and
${\cal B}_c^{\rm 2d\,Potts}=1.16(1)$ \cite{Tom_eacute__2002}.

\section{Localization of staggered modes: numerical results}
\label{sec:numloc}

We numerically simulated pure $\mathbb{Z}_3$ gauge theory on 2+1
dimensional $N_s^2\times N_t$ lattices, with $N_t=4$ and
$N_s=20,24,28,32$, and with $\beta$ values on both sides of the
deconfinement transition, using a standard Metropolis algorithm. For
each $\beta$ and $N_s$ we collected 1500 well decorrelated gauge
configurations, and for each of them we computed the full set of
eigenvalues and eigenvectors of the staggered Dirac operator,
Eq.~\eqref{eq:loc1}, using the LAPACK
library~\cite{anderson1999lapack}. We also measured the local
plaquettes and Polyakov loops in order to study their correlation with
the staggered eigenmodes.

As explained in Section \ref{sec:loc}, in the deconfined phase
($\beta>\beta_c$) averages are computed separately for configurations
in the physical, real Polyakov-loop sector ($\Im \bar{P} =0$,
$\Re \bar{P} >0$) and in the complex sectors ($\Im \bar{P} \neq 0$,
$\Re \bar{P} <0$). Since the two complex sectors
$\Im \bar{P} \gtrless 0$ yield identical results thanks to
$C$-invariance, it suffices to study the sector $\Im \bar{P} > 0$. To
ensure that configurations are in the desired center sector we make a
suitable center transformation, Eq.~\eqref{eq:z3_9}, whenever needed,
thus effectively collecting 1500 configurations in both sectors under
study. In the confined phase ($\beta<\beta_c$) where center symmetry
is realized, averages are instead computed over the full set of
configurations, covering evenly all the sectors.

For a finite ensemble of gauge configurations, the local averages
$\overline{\Oc}(\lambda,V)$, Eq.~\eqref{eq:loc3}, are estimated by
averaging over the modes in small disjoint intervals of equal width
$\Delta\lambda=0.05$ (in lattice units), and assigning the result to
the average eigenvalue in the bin. Errors are estimated via the
standard jackknife method. Additional care is required for the
calculation of the PR in the presence of degenerate eigenvalues, which
do show up in discrete gauge theories in small and moderate volumes
(see Ref.~\cite{Baranka:2021san} for the case of $\mathbb{Z}_2$).  In
this case a single value of the PR is assigned to the whole degenerate
subspace by means of a suitable average, and included in the bin
average with multiplicity equal to the dimension of the subspace (see
the appendix of Ref.~\cite{Baranka:2021san} for details). The fractal
dimension is then estimated from $\overline{\textrm{PR}}(\lambda,V)$
using pairs of system spatial sizes $N_{s_{1,2}}$ via
\begin{equation}
  \label{eq:fdim_est}
  \alpha(\lambda;N_{s_1},N_{s_2}) = 2 + \f{\log
    \left[\overline{\textrm{PR}}(\lambda,N_{s_1}^2)/
      \overline{\textrm{PR}}(\lambda,N_{s_2}^2)\right]}{\log
    (N_{s_1}/N_{s_2})}\,. 
\end{equation}
The corresponding error is obtained by standard linear propagation.
For sufficiently large volumes where eigenvalues become dense,
unfolded spacings can be computed by dividing the level spacings
$\lambda_{i+1}-\lambda_i$ by the average level spacing
$({\cal V}\rho)^{-1}$, i.e.,
\begin{equation}
  \label{eq:unf_def}
  \begin{aligned}
    s_i=x_{i+1}-x_i&={\cal V}
    \int_{\lambda_i}^{\lambda_{i+1}}d\lambda\,
    \rho(\lambda)\\
    &\simeq (\lambda_{i+1}-\lambda_{i}){\cal V}\rho(\lambda_i) \,.
  \end{aligned}
\end{equation}
For a finite ensemble, this is done in practice by estimating the
average spacing in each spectral bin by including those spacings
$\lambda_{l+1}-\lambda_{l}$ for which $\lambda_l$ lies in the bin, and
dividing $\lambda_{i+1}-\lambda_{i}$ by the average spacing in the bin
where $\lambda_{i}$ belongs. Other definitions are possible (e.g., one
could include in each bin average only those spacings for which the
middle point between the eigenvalues lies in the bin), but they are
all equivalent in the large-volume limit. This practical definition
of unfolded spacings avoids problems with accidental degeneracies of
eigenvalues.

In the deconfined phase the Polyakov loops become spatially ordered,
breaking spontaneously the center symmetry of the system, and inducing
strong correlations among different time slices. In a first
approximation, typical gauge configurations can be thought of as
fluctuating around the perfectly ordered configuration
$P(\vec{x})= e^{i \phi_0}$ with spatial links all equal to unity. A
fermion in this gauge background is equivalent to a free fermion
subject to nontrivial temporal boundary conditions, a setup for which
the staggered operator can be diagonalized exactly.  The positive
eigenvalues read
\begin{equation}
  \label{eq:free_stag_phi}
  \lambda_{k,j_1,j_2} = \sqrt{(\sin\omega_k)^2+(\sin p_{j_1})^2+(\sin p_{j_2})^2}\,,
\end{equation}
where $\omega_k=\f{\phi_0+(2k+1)\pi}{N_t}$ with $k= 0,\ldots,N_t-1$,
and $p_{j_{1,2}}=\frac{2 \pi j_{1,2}}{N_s}$ with
$j_{1,2}=0,\ldots, N_s-1$.  The free spectrum
Eq.~\eqref{eq:free_stag_phi} lies in the interval
$[\lambda_{\rm L},\lambda_{\rm H}]$ where
\begin{equation}
  \label{eq:spec_inter}
  \lambda_{\rm L}=\min_k|\sin\omega_k|\,,
  \quad \lambda_{\rm H} =\sqrt{\max_k\, (\sin\omega_k)^2+2} \,.
\end{equation}
When looking at the results for the interacting spectrum in the
deconfined phase, one can use the endpoints of the free spectrum for
the appropriate values of $\phi_0$ and $N_t$ to separate the bulk from
the low modes ($\lambda <\lambda_{\rm L}$) and the high modes
($\lambda>\lambda_{\rm H}$). These points are marked by vertical
dashed lines in our figures. For $N_t = 4$, in the real sector
($\phi_0=0$) one has $\lambda_{\rm L}^{(r)}=\frac{1}{\sqrt{2}}$ and
$\lambda_{\rm H}^{(r)}=\sqrt{\frac{5}{2}}$, while in the complex
sectors ($\phi_0=\pm \frac{2 \pi}{3}$) one has
$\lambda_{\rm L}^{(c)}=\sin\f{\pi}{12}$ and
$\lambda_{\rm H}^{(c)}=\sqrt{(\sin\f{5 \pi}{12})^2+2}$. Since in the
confined phase all center sectors contribute, in the corresponding
figures all the special points mentioned above are marked by vertical
dashed lines. The square root of the average eigenvalue squared,
$\lambda_*=\sqrt{\frac{3}{2}}$, is also marked by a solid vertical
line in our figures. This point in the spectrum is characterized by a
sizeable degeneracy of eigenmodes for the medium-size lattices used
here, and often corresponds to a noticeable dip or peak in the plots
of the various observables.

\subsection{Mode size, fractal dimension, and spectral statistics}
\label{sec:numloc_modesize}

\begin{figure}[t]

  \includegraphics[width=0.48\textwidth]{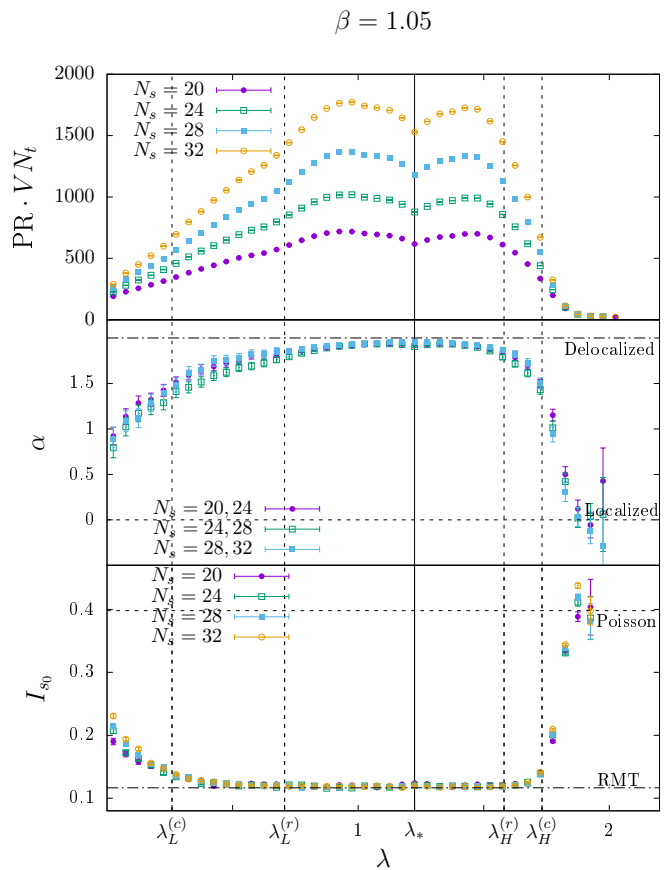}
  \caption{Eigenmode properties in the confined phase. Top: Mode size.
    Center: Fractal dimension. Horizontal lines mark expectations for
    localized (dashed) and fully delocalized modes (dot-dashed).
    Bottom: Integrated unfolded level spacing distribution. Horizontal
    lines mark expectations for Poisson (dashed) and RMT statistics
    (dot-dashed). Here $N_t=4$.}
  \label{fig:all_conf}
\end{figure}

\begin{figure}[t]
  \includegraphics[width=0.48\textwidth]{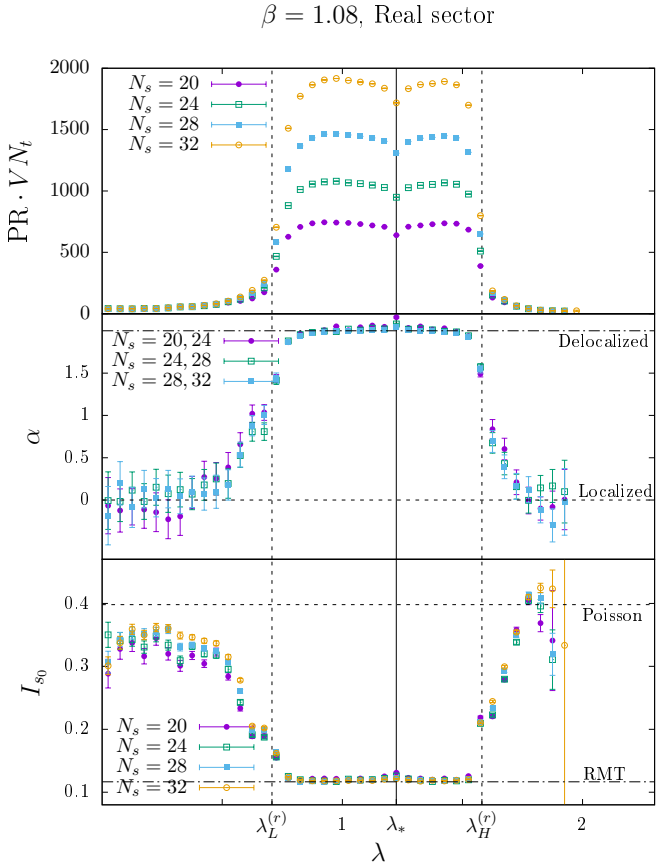}
  \caption{Eigenmode properties in the deconfined phase -- real
    sector. Top: Mode size.  Center: Fractal dimension. Horizontal
    lines mark expectations for localized (dashed) and fully
    delocalized modes (dot-dashed). Bottom: Integrated unfolded level
    spacing distribution. Horizontal lines mark expectations for
    Poisson (dashed) and RMT statistics (dot-dashed). Here $N_t=4$.}
  \label{fig:all_dec_ph}
\end{figure}

\begin{figure}[t]
  \includegraphics[width=0.48\textwidth]{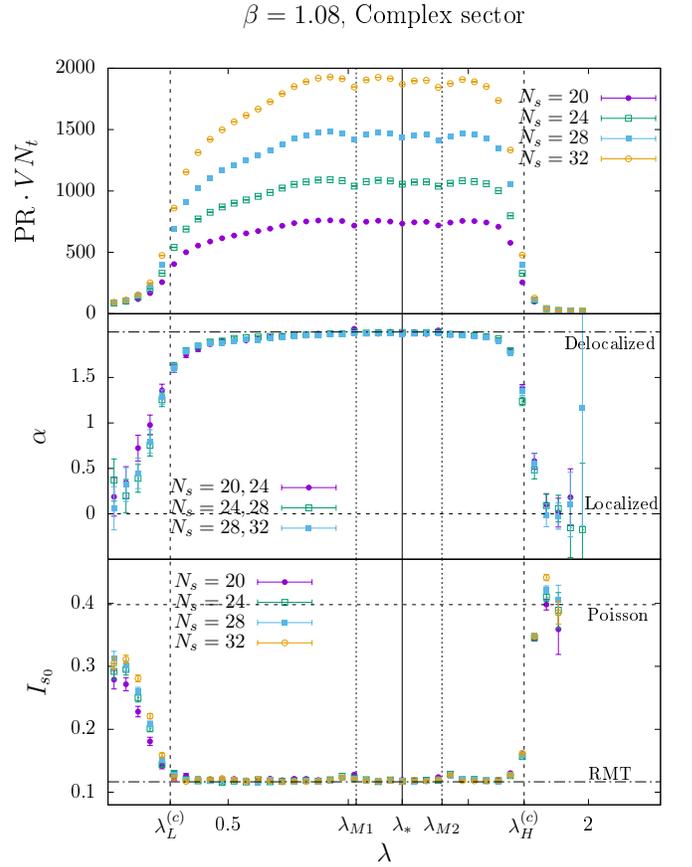}
  \caption{Eigenmode properties in the deconfined phase -- complex
    sectors. Top: Mode size.  Center: Fractal dimension. Horizontal
    lines mark expectations for localized (dashed) and fully
    delocalized modes (dot-dashed). Bottom: Integrated unfolded level
    spacing distribution. Horizontal lines mark expectations for
    Poisson (dashed) and RMT statistics (dot-dashed). Here $N_t=4$.}
  \label{fig:all_dec_unph}
\end{figure}

As explained in Section \ref{sec:loc}, localization is conveniently
detected by studying the size and the fractal dimension of the
eigenmodes, and the statistical properties of the corresponding
eigenvalues. Numerical results for the average mode size
$N_t V \cdot \overline{\textrm{PR}}(\lambda,V) =
\overline{\textrm{IPR}^{-1}}(\lambda,V) \sim V^{\alpha(\lambda)}$ and
the corresponding fractal dimension $\alpha(\lambda)$,
Eq.~\eqref{eq:loc3ter}, and for the local average of the integrated
unfolded level spacing distribution $I_{s_0}(\lambda,V)$,
Eq.~\eqref{eq:loc_stat4}, are shown in
Figs.~\ref{fig:all_conf}--\ref{fig:all_dec_unph} for typical $\beta$
values both below and above $\beta_c(N_t=4)$ (see
Tab.~\ref{tab:potts}), and for different lattice sizes.

\subsubsection{Confined phase}
\label{sec:numloc_modesize_conf}

Results for the confined phase are shown in Fig.~\ref{fig:all_conf}.
In this phase the low modes are delocalized, but with a nontrivial
fractal dimension close to 1.  A similar behavior was found for gauge
group $\mathbb{Z}_2$~\cite{Baranka:2021san}.  In the standard language
of disordered systems, these modes are therefore critical.  As
$\lambda$ increases, modes become more and more delocalized, with an
increasing fractal dimension which gets close to 2 as one enters the
bulk of the spectrum.  Our estimates of $\alpha$ remain always
slightly smaller than 2, but this may as well be only a finite size
effect, with full delocalization eventually reached in the middle of
the spectrum for larger volumes. Finally, at the high end of the
spectrum the fractal dimension is compatible with zero, indicating
that modes are localized.

Results for $I_{s_0}$ (Fig.~\ref{fig:all_conf}, bottom) support this
picture, with low modes displaying a nontrivial, $\lambda$-dependent
statistics intermediate between Poisson and RMT, and only mildly
dependent on the volume; bulk modes compatible with RMT behavior,
except towards both ends of the bulk where they depart from it; and
high modes quickly becoming compatible with Poisson statistics. We
then expect an extend region of critical modes at the low end of the
spectrum, either extending all the way across the bulk or containing a
region of fully extended modes; and a region of localised modes at the
high end of the spectrum. Which of these alternatives is realized for
the bulk modes cannot be decided with the current lattice and gauge
ensemble sizes.

\subsubsection{Deconfined phase}
\label{sec:numloc_modesize_dec}

\paragraph{Real sector}

Results for the real sector in the deconfined phase are shown in
Fig.~\ref{fig:all_dec_ph}.  There both low and high modes are clearly
localized.  Bulk modes instead are delocalized, with a fractal
dimension quickly rising from around 1 to 2 (i.e., full
delocalization) as one enters the bulk from either end. In the
transition regions $\alpha$ has little to no volume dependence while
being clearly separated from 2. This is the kind of behavior expected
for a BKT-type Anderson transition (see Section \ref{sec:loc_AT}).
While difficult to determine precisely with the available data, the
mobility edges are found in the vicinity of $\lambda_{\rm L}^{(r)}$
and $\lambda_{\rm H}^{(r)}$.

Results for the spectral statistics (Fig.~\ref{fig:all_dec_ph},
bottom) support the picture obtained from the mode size. In the
low-mode region $I_{s_0}$ is rather flat near zero, and closer to the
value expected for Poisson statistics than that expected for RMT
statistics. More importantly, it shows a tendency to become flatter
and closer to the Poisson expectation as the volume is increased.
Although the volume dependence is not much stronger than in the
confined phase, the relative flatness of $I_{s_0}$ at the low end of
the spectrum in the deconfined phase suggests that low modes all share
the same spectral statistics, as opposed to the changing statistics
observed in the confined phase. For bulk and high modes one finds
spectral statistics very close respectively to RMT and to Poisson
statistics, exactly as in the confined phase.  Near the mobility edges
one finds for $I_{s_0}$ a value intermediate between the Poisson and
the RMT expectation, and approximately volume independent, again
supporting the expectation that the Anderson transitions are of BKT
type.

\paragraph{Complex sectors}

Results for the complex sectors are shown in
Fig.~\ref{fig:all_dec_unph}.  Also in these sectors the low modes
appear to be localized, with their fractal dimension
(Fig.~\ref{fig:all_dec_unph}, center) tending to zero as the lattice
sizes used for its estimate are increased.  As discussed in Sections
\ref{sec:intro} and \ref{sec:newsi}, this cannot be explained in terms
of ``energetically'' favorable islands of Polyakov-loop fluctuations
alone. Notice that the mode size of the low modes is larger in the
complex sectors than in the real sector. For bulk and high modes the
situation is the same found in the real sector, i.e., bulk modes are
delocalized (and fully so deep in the bulk) and high modes are
localized, with mobility edges again in the vicinity of $\lambda_{\rm
L}^{(c)}$ and $\lambda_{\rm H}^{(c)}$, therefore different from those
found in the real sector and closer to the spectrum edges.

In Fig.~\ref{fig:all_dec_unph}, and in all other plots concerning the
complex sectors in the deconfined phase, we also show two other
special points in the spectrum, $\lambda_{\rm M1}$ and
$\lambda_{\rm M2}$, marked by vertical dotted lines.  These correspond
to choosing $j_{1,2}$ so that $(\sin p_{j_1})^2+(\sin p_{j_2})^2=1$,
and $k=0$ or $1$, corresponding to $\sin\omega_0=\sin\f{\pi}{12}$ or
$\sin\omega_1=\sin\f{5\pi}{12}$, in the free spectrum
Eq.~\eqref{eq:free_stag_phi}. At these points we also found a sizeable
degeneracy of eigenmodes, as well as dips or peaks in the various
observables. Remarkably, Fig.~\ref{fig:all_dec_unph}, center, shows
that in the interval between these two points the fractal dimension of
bulk modes equals 2 within errors, departing from it outside this
interval and dropping towards zero as one enters the low or high mode
region. Moreover, in the whole bulk region there is little to no
dependence on the lattice volumes used for the estimate of
$\alpha$. This suggests that also in the complex sectors the Anderson
transition at the mobility edges is of BKT type.

This picture is again supported by our findings for the spectral
statistics (Fig.~\ref{fig:all_dec_unph}, bottom), which are similar to
those obtained in the real sector.  The main difference, besides the
position of the mobility edges, is the lower value of $I_{s_0}$
attained by the low modes in the complex sectors on the available
volumes. This agrees with the fact that in the complex sectors the
localized low modes are more extended than in the real sector (see
Figs.~\ref{fig:all_dec_ph} and \ref{fig:all_dec_unph}, top panels). In
fact, the typical overlap between two distinct, well localized modes
in a finite system is small but nonetheless finite, vanishing only in
the infinite-volume limit where they can be arbitrarily far apart.
This leads to a finite correlation between the fluctuations of the
corresponding eigenvalues under a change in the gauge field
configuration, and so to a deviation from Poisson statistics, which in
particular shows up as a finite-volume effect in the unfolded level
spacing distribution. Clearly, localized modes of larger size have
larger typical overlap, and so a larger deviation from Poisson
statistics.

\begin{figure}[t]
  \includegraphics[width=0.48\textwidth]{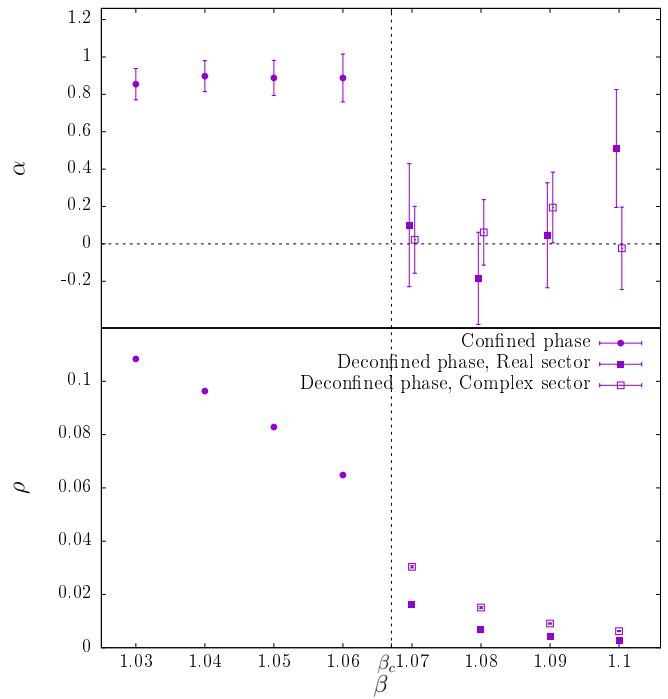}
  \caption{Top: fractal dimension of near-zero modes as a function of
    $\beta$. Data points for the real and complex sectors in the
    deconfined phase have been symmetrically shifted horizontally for
    clarity. Bottom: spectral density of near-zero modes as a function
    of $\beta$. Here $N_s=32$.}
  \label{fig:fdim_T}
\end{figure}

\paragraph{Near-zero modes}

The localization properties of the near-zero modes as a function of
$\beta$ are summarized in Fig.~\ref{fig:fdim_T} (top), where we show
the fractal dimension $\alpha$ of modes in the lowest spectral bin
$\lambda_n \in [0,\Delta\lambda]$. Central values correspond to
$\alpha$ obtained from the pair of largest sizes $(28,32)$. Error bars
are obtained by adding in quadrature the statistical error and the
systematic error due to finite-size effects, estimated as the standard
deviation of the sample of values of $\alpha$ obtained from all
possible pairs of sizes $(N_{s1},N_{s2})$.

A drastic change takes place at the deconfinement transition, both in
the real and in the complex sectors, as modes suddenly turn from
critical to localized, with a fractal dimension compatible with zero
within numerical errors. For the real sector this gives further
support to the general expectation that localized low modes appear
right at the deconfinement transition. For the complex sector this
requires instead to revisit the sea-islands picture. Error bars are
noticeably larger in the deconfined phase than in the confined phase,
both close to $\beta_c$ and high above it. Close to the transition,
this is mostly due to large finite-size effects caused by the
correlation length of the system diverging at $\beta_c$, causing also
a visible increase in the error for the closest point in the confined
phase. Another source of uncertainty is the low count of near-zero
modes in the deconfined phase, discussed below.

In Fig.~\ref{fig:fdim_T} (bottom) we show the spectral density of
near-zero modes, i.e., the average number of modes per unit volume in
the lowest spectral bin divided by $\Delta\lambda$. This decreases
with increasing $\beta$, changing more rapidly near the transition,
and becoming very small although still nonzero at large $\beta$.  This
explains why the error bars for $\alpha$ remain large also far from
the transition.

Different values are found in the two center sectors, with a lower
density in the real one. These findings are consistent with the quite
general pattern of deconfinement improving on the chiral symmetry
properties of the system, indicated here by the large decrease in the
near-zero spectral density; and with the fact that fermions prefer the
real Polyakov-loop sector over the complex ones.

\subsection{Gauge observables} 
\label{sec:numloc_gauge_obs}

\begin{figure}[t]
  \includegraphics[width=0.48\textwidth]{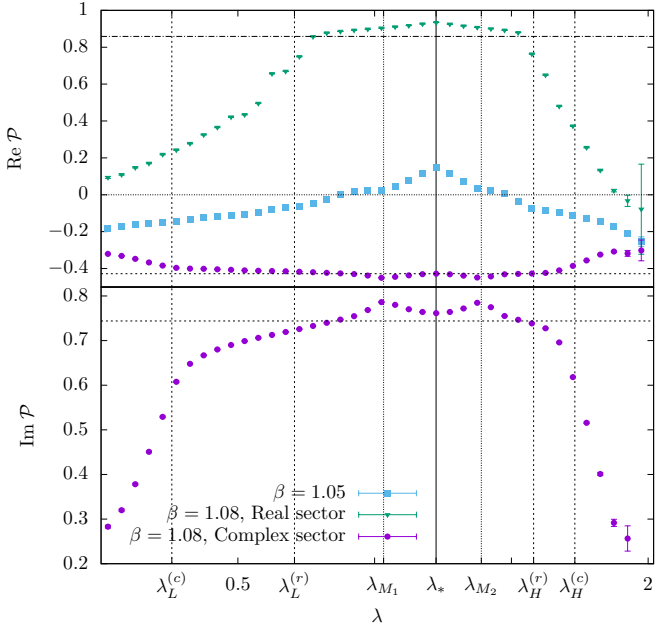}
  \caption{The real part of the Polyakov loop weighted by the modes
    (top), and its imaginary part in the deconfined phase in the
    complex sector $\Im\bar{P}>0$ (bottom). The horizontal lines
    correspond to the real or the imaginary part of the average
    Polyakov loop $\la P\ra$, as appropriate, in the deconfined phase --
    real sector (dot-dashed), confined (short dashed), and deconfined
    phase -- complex sector (long dashed). Here $N_s=32$.}
  \label{fig:sea1}
\end{figure}

In Fig.~\ref{fig:sea1} we show our results for the correlation between
eigenmodes and Polyakov loops.  For bulk modes, $\Re\overline{\cal P}$
is always close to the real part of the expectation value
$\la \Re P\ra$ of the Polyakov loop.  In the confined phase
($\la P\ra=0$) and in the deconfined phase in the real sector
($\langle \Re P \rangle > 0 $) there is a mild correlation with real
Polyakov loops, while in the deconfined phase in the complex sectors
($\langle \Re P \rangle < 0 $) the deviation from $\la \Re P\ra$ is
minimal. For low and high modes, the behavior of
$\Re\overline{\cal P}$ is quite different in the three cases. In the
confined phase $\Re\overline{\cal P}$ shows a mild correlation of the
eigenmodes with complex Polyakov loops. In the deconfined phase in the
real sector $\Re\overline{\cal P}$ shows a strong correlation of the
eigenmodes with complex Polyakov loops, with $\Re\overline{\cal P}$
reaching down to almost 0 for low modes, and to negative values for
high modes. This means that around 2/3 or more of the weight of the
low modes is found on islands of Polyakov-loop fluctuations. This
agrees with the standard sea-islands picture, as there are very few
islands of fluctuations and still a large fraction of the mode is
localized on those islands.

\begin{figure}[t]
  \includegraphics[width=0.48\textwidth]{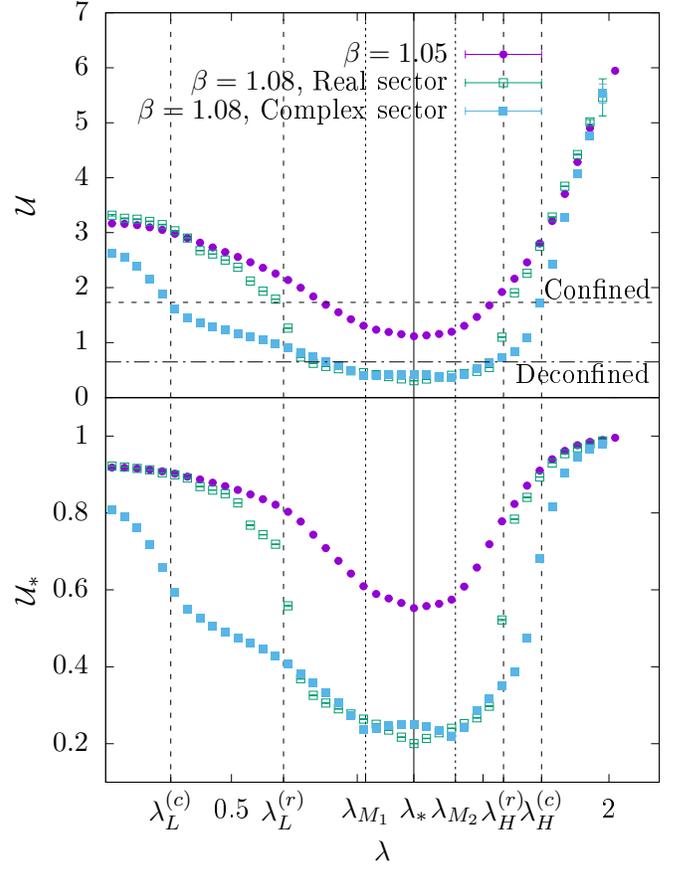}
  \caption{Average number of non-trivial plaquettes touched by a mode
    (top), and weight of modes on negative plaquettes
    (bottom). Horizontal lines in the top panel correspond to
    $8 \la 1-U_{\mu\nu}\ra$ at the given $\beta$ values in the
    confined and deconfined phase. Here $N_s=32$.}
  \label{fig:negplaq}
\end{figure}

\begin{figure*}[t]
  \includegraphics[width=0.97\textwidth]{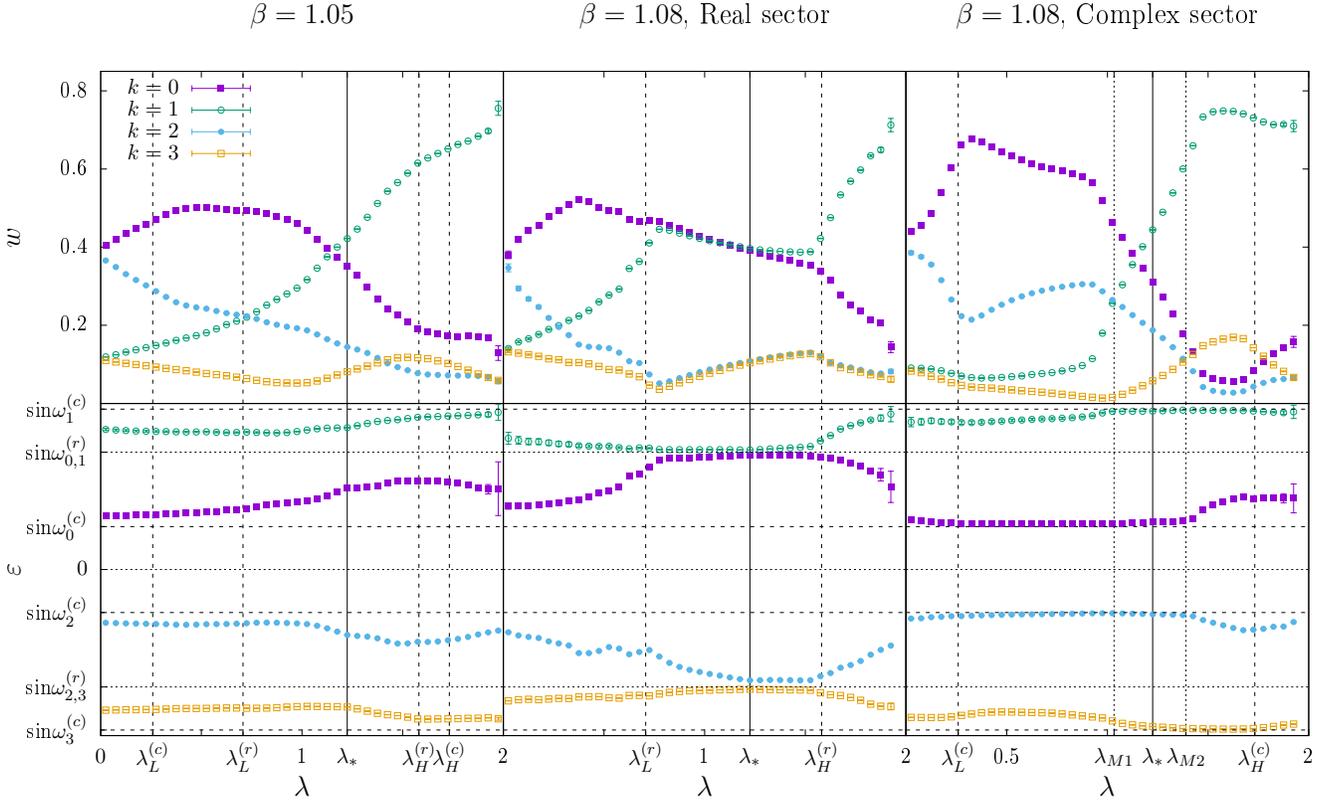}
  \caption{Standard sea-islands picture: average weight (top row) and
    energy (bottom row) per branch, in the confined phase (left
    panels), deconfined phase -- real sector (center panels), and
    deconfined phase -- complex sectors (right panels). Here $N_s=20$
    and $N_t=4$.}
  \label{fig:si1eta}
\end{figure*}

On the other hand, in the deconfined phase in the complex sectors
$\Im\overline{\cal P}$ shows a strong correlation of these modes with
Polyakov-loop fluctuations in the opposite complex sector, and
moreover $\Re\overline{\cal P}$ shows a mild but sizeable correlation
with real Polyakov-loop fluctuations. This means that the localized
low and high modes have larger weight on Polyakov-loop fluctuations
than delocalized modes, including on the ``energetically unfavorable''
real Polyakov-loop fluctuations, which for the low modes contradicts
the general expectations of the standard sea-islands picture.

In Fig.~\ref{fig:negplaq} we show our results for the correlation
between eigenmodes and nontrivial plaquettes. Qualitatively, the
situation is the same found for $\mathbb{Z}_2$ gauge
theory~\cite{Baranka:2021san}. Low and high modes always show a strong
correlation with nontrivial plaquettes, as signalled by a value of
$\overline{{\cal U}_*}$ close to 1; and with clusters of nontrivial
plaquettes in particular, as signalled by a value of
$\overline{\cal U}$ larger than 1. For the low modes this happens
independently of their localization properties, although in the
deconfined phase, where nontrivial plaquettes become less frequent,
this indicates localization. For modes deep in the bulk (near
$\lambda_*$) one finds instead a value of $\overline{\cal U}$ close to
(and below) the value $8 \la 1-U_{\mu\nu}\ra$ expected for perfectly
delocalized modes ($|\psi|^2 = 1/(N_t V)$).  The correlation with
clusters of nontrivial plaquettes generally increases as one moves
away from the deep bulk near $\lambda_*$, where modes are repelled by
them. The mode weight on negative plaquettes keeps similarly
increasing as one moves away from $\lambda_*$. Notably, in the
deconfined phase the special points in the spectrum correspond to
clear changes in the behavior of $\overline{\cal U}$ and
$\overline{{\cal U}_*}$.

\begin{figure*}[ht]
  \includegraphics[width=0.48\textwidth]{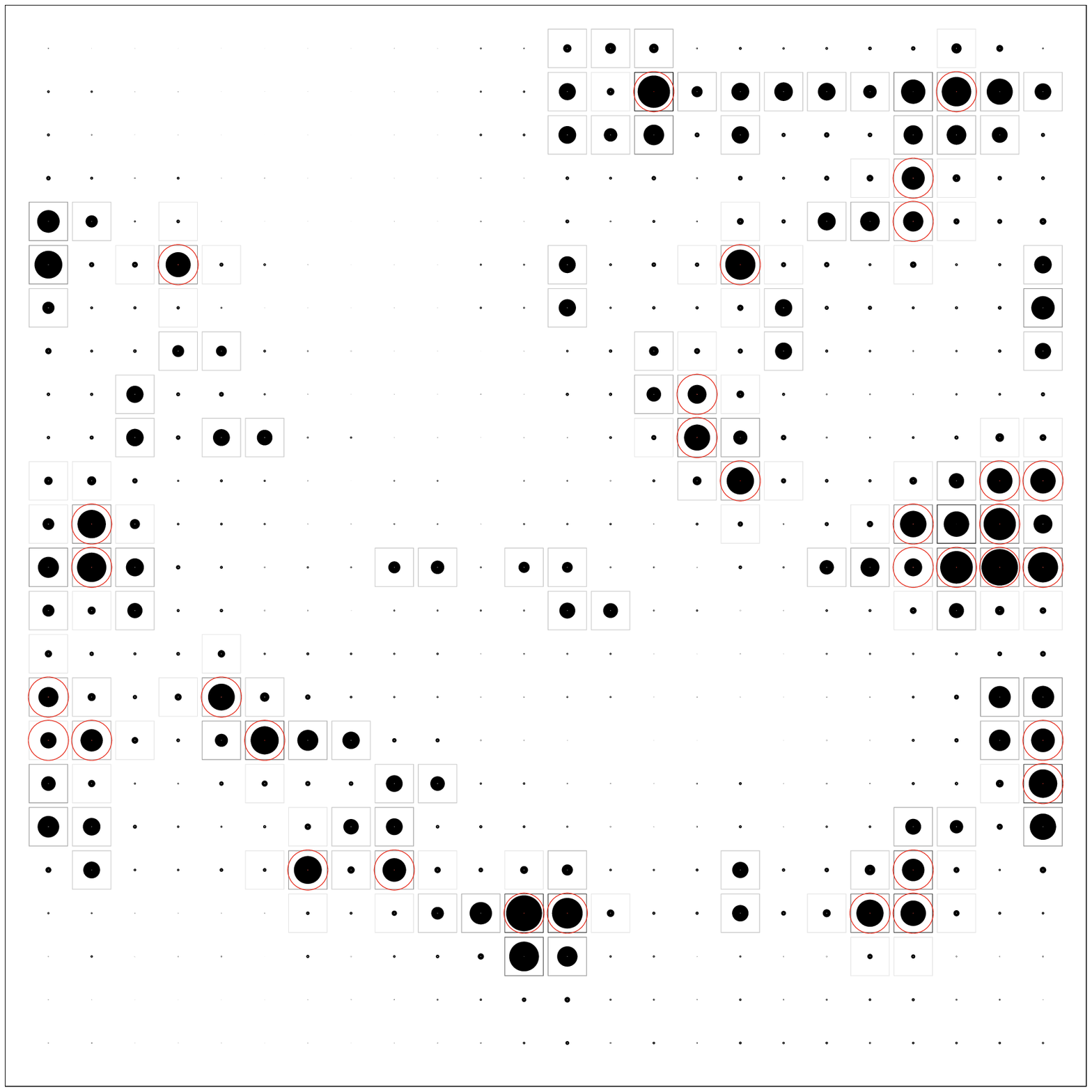}\hfil\includegraphics[width=0.48\textwidth]{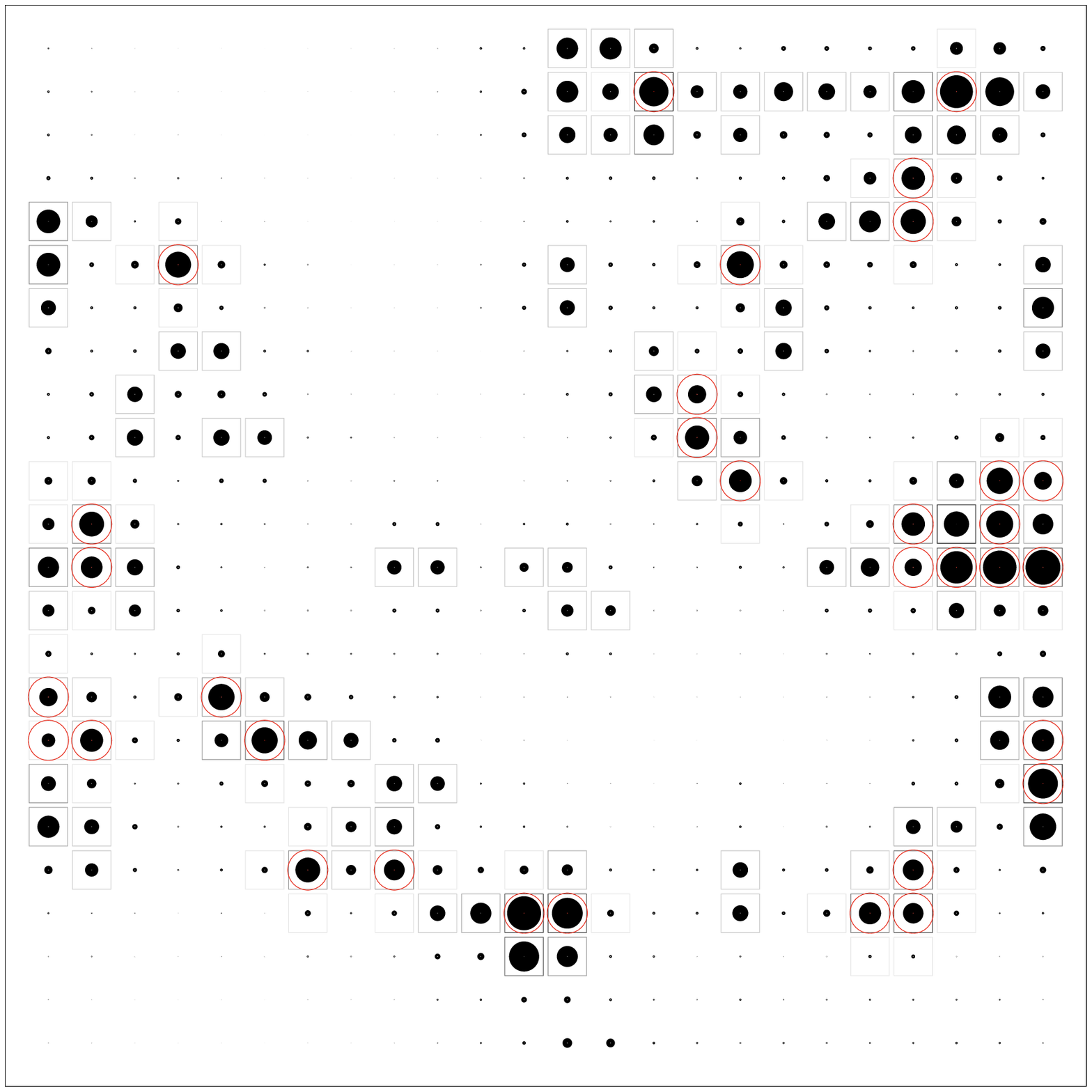}

  \vspace{0.2cm}
  \includegraphics[width=0.48\textwidth]{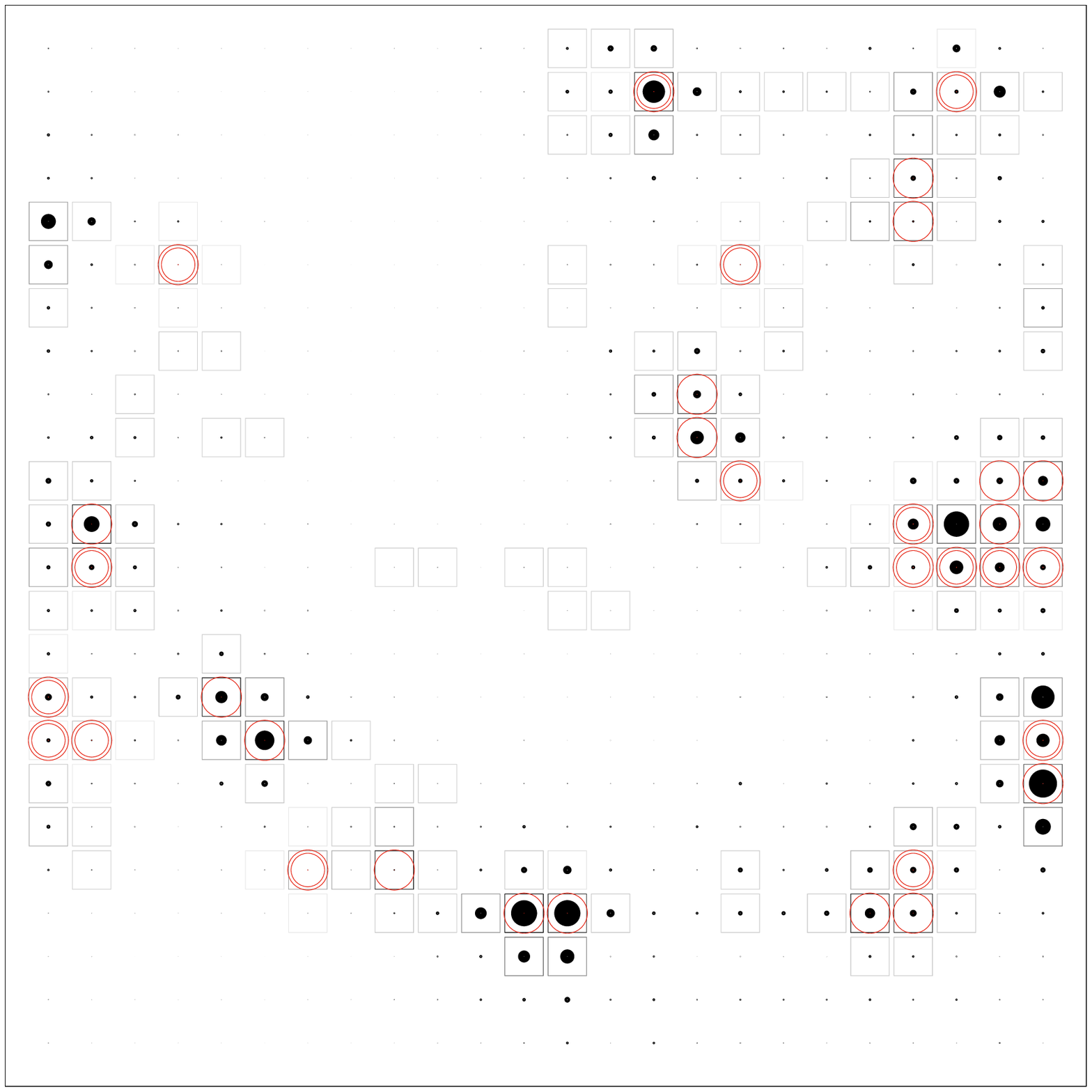}\hfil\includegraphics[width=0.48\textwidth]{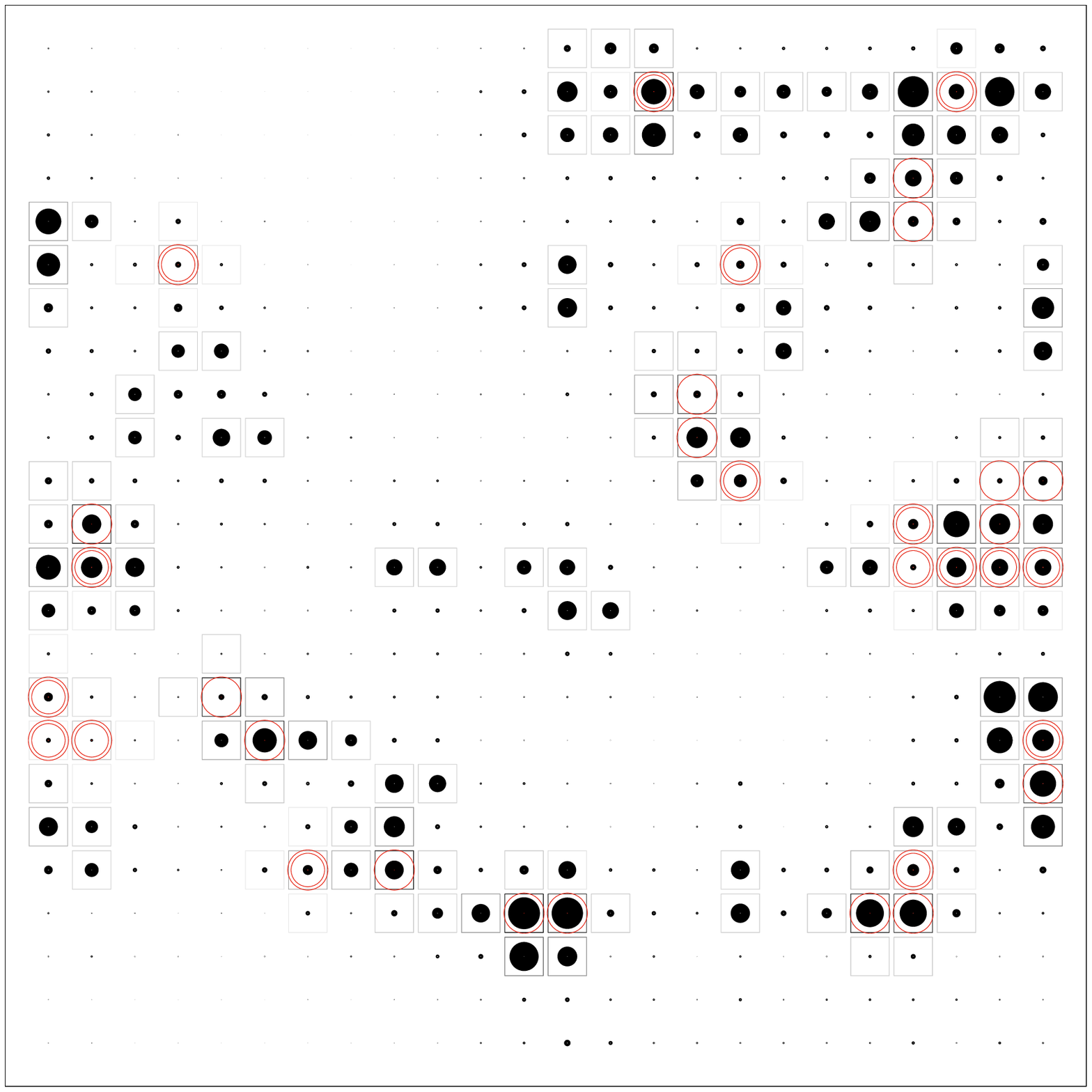}
  \caption{Correlation between mode amplitude and hopping terms for a
    single $32^2\times 4$ gauge configuration at $\beta=1.08$ (top
    row), and for its center-rotated version in the complex sector
    ${\rm Im}\, \bar{P}> 0$ (bottom row), for low modes
    ($\lambda_n\le \lambda_L$, left panels) and high modes
    ($\lambda_n\ge \lambda_H$, right panels). Dots are located on
    spatial lattice sites, and squares cover the corresponding
    Wigner-Seitz cell.  The dot size is proportional to the low and
    high mode amplitudes $p_L$ and $p_H$, Eq.~\eqref{eq:moddens}, with
    an extra enhancement factor $5/3$ in the complex sector for better
    visualization. Red circles denote nontrivial Polyakov loops; in
    the complex sector, a double circle denotes a real Polyakov
    loop. A darker shade of gray of the squares corresponds to a
    larger ${\cal A}$, Eq.~\eqref{eq:newsi_meas}, indicating a more
    favorable place for localization according to the refined
    sea-islands picture.}
  \label{fig:newsi_conf}
\end{figure*}

\begin{figure*}[th]
  \includegraphics[width=0.48\textwidth]{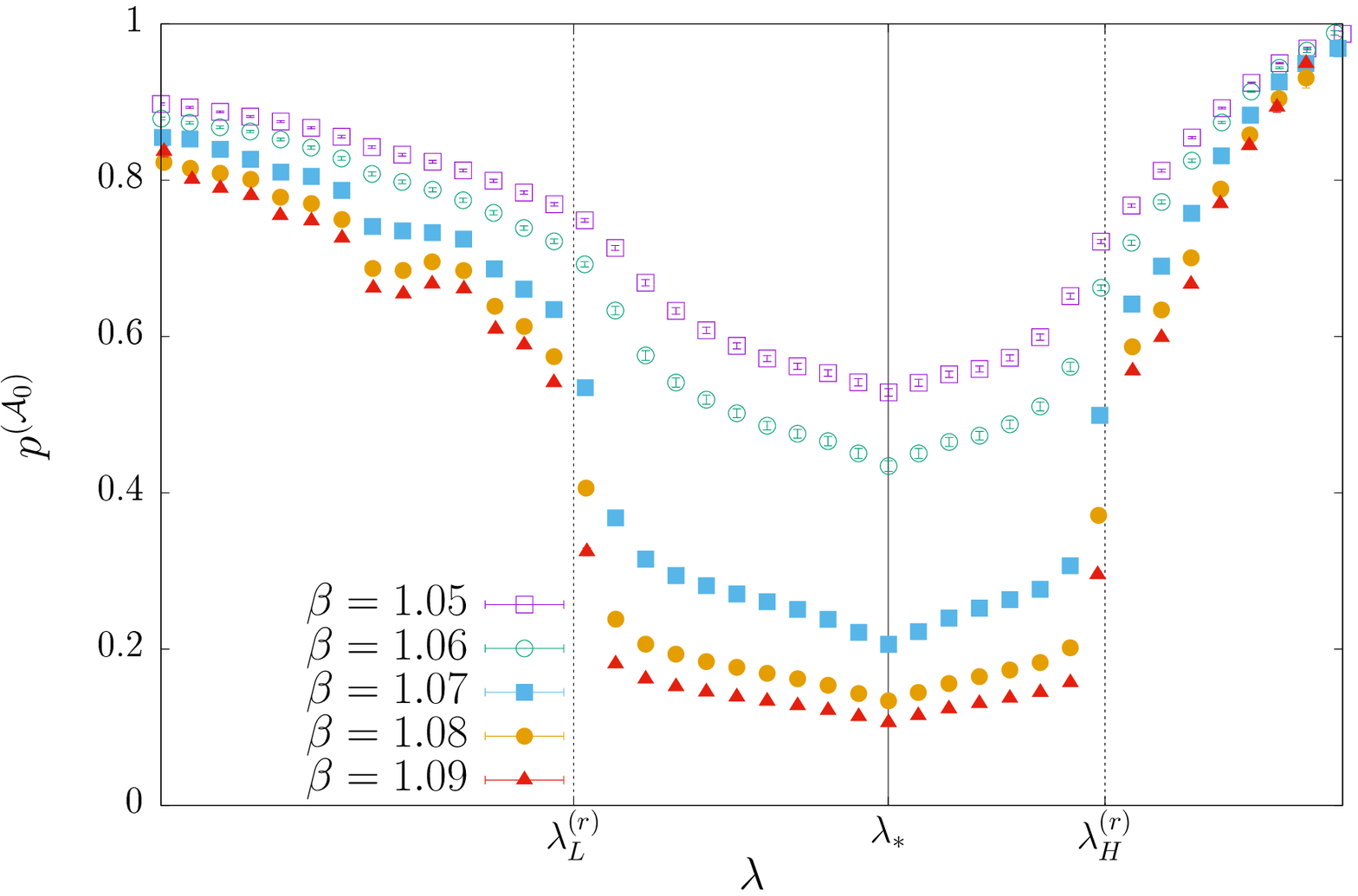}\hfil\hfil
  \includegraphics[width=0.48\textwidth]{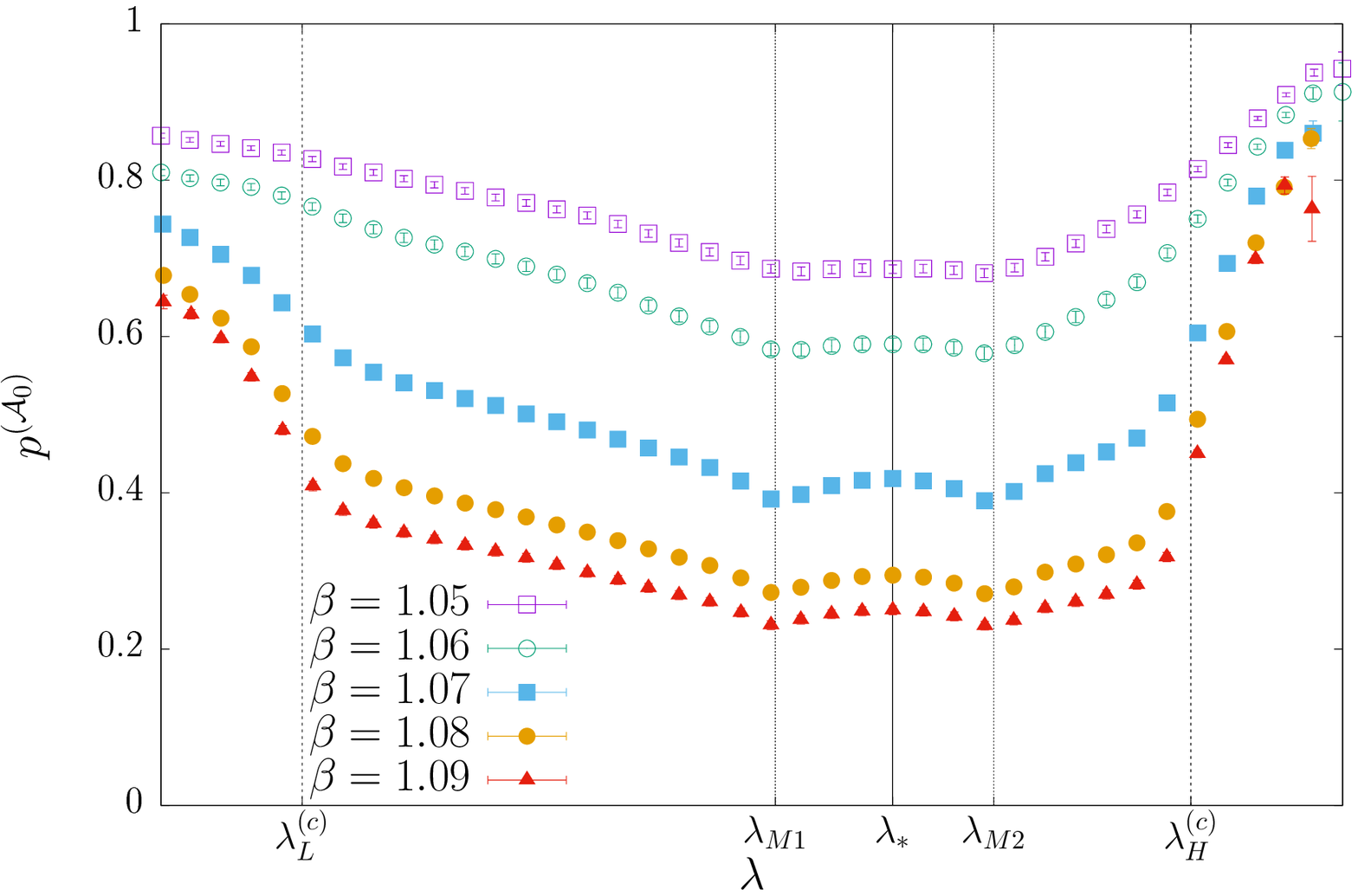}
  \caption{Average mode weight on sites where
    ${\cal A}(\vec{x})\ge {\cal A}_0$, in the real sector (left) and
    in the complex sectors (right), on a $32^2\times 4$ lattice. Here
    ${\cal A}_0=1/16$.}
  \label{fig:newsi_ave}
\end{figure*}

\subsection{Standard sea-islands picture}
\label{sec:numloc_si}

To study the standard sea-islands picture in detail, we measured the
weights and average energy of the various branches of the eigenmodes,
Eqs.~\eqref{eq:DA9_alt} and \eqref{eq:DA10_alt}, averaged locally in
the spectrum, Eq.~\eqref{eq:loc3}. For the confined phase we chose
$\beta=1.05$, while for the deconfined phase we chose $\beta=1.08$ and
looked at the real sector and at the complex sector with
$\Im\bar{P}>0$. For each setup we used 100 configurations on a
$20^2\times 4$ lattice.  Our results are shown in
Fig.~\ref{fig:si1eta}. Irrespectively of the phase or center sector,
the lowest positive modes have the largest weight on the $k=0$ branch,
as expected, but also a sizeable weight on the corresponding negative
branch $k=\f{N_t}{2}=2$, up to (and also partially including, for the
confined phase and the complex sectors in the deconfined phase) the
bulk region. In the deconfined phase in the real sector, $k=0$ and
$k=1$ have practically equal weights throughout the bulk. This can be
understood by noticing that these two branches are degenerate for
$\phi(\vec{x})=0$, so that for bulk modes, delocalized all over the
sea of ordered and trivial Polyakov loops, one expects that they fully
mix. This kind of degeneracy between branches is generally expected in
the real sector, since for $\phi(\vec{x})=0$ one finds
$\sin\tilde\omega_{\f{N_t}{2}-1-k}(0)= \sin\tilde\omega_{k}(0)$. The
same argument applies to the pair of branches $k=1,3$. The branch
$k=N_t-1=3$ never contributes substantially to the eigenmodes, except
at the high end of the bulk in the deconfined phase in the complex
sectors, where it gives the second largest contribution.

Except for the large mixing of branches in the bulk of the spectrum
observed in the deconfined phase in the real sector, whose origin is
clear, our results show that in the bulk and in the high-mode regions
there is always one of the coupled Anderson models dominating the wave
function (except of course in the transition regions where the
dominant Anderson model changes). For the lowest modes, instead, the
$k=0$ (positive) and the $k=\f{N_t}{2}=2$ (negative) branch contribute
comparably. This requires only a rather mild adjustment to the
expectations of the standard sea-islands picture, with the positive
and the negative energy level closest to zero both playing a
role. Nonetheless, this does not help explaining why low modes
localize in the complex Polyakov-loops sectors.

Results for the average energy level seen by a branch are again in
nice agreement with the standard expectations in the case of the real
sector of the deconfined phase, with the leading $k=0$ and $k=2$
branches of the low modes clearly showing a preference for
Polyakov-loop fluctuations to the complex sectors. For bulk modes
instead one observes an approximate degeneracy of the $k=0,1$ and of
the $k=2,3$ branches, as one would expect based on the discussion
above. In the complex sectors of the deconfined phase, instead, the
low localized modes show a small but clear deviation from what one
would na\"ively expect, indicating again that they are surprisingly
favoring real Polyakov-loop fluctuations over the more
``energetically'' convenient sea. This further calls for reconsidering
the standard sea-islands picture.

\subsection{Refined sea-islands picture}
\label{sec:numloc_nsi}

To test the refined sea-islands picture discussed in
Sec.~\ref{sec:newsi}, we have directly inspected a few gauge
configurations in the deconfined phase, both in the real and in the
complex Polyakov-loop sectors, looking for correlations between where
localized modes live and the locations where the hopping terms deviate
from $A_j\approx \mathbf{0}$. As a measure of this deviation we used
\begin{equation}
  \label{eq:newsi_meas}
  {\cal A}(\vec{x}) = \frac{1}{8}\sum_{j=1}^2 \tr\left\{
    A_j(\vec{x}){}^\dag A_j(\vec{x}) + 
    A_j(\vec{x}-\hat\jmath){}^\dag
    A_j(\vec{x}-\hat\jmath) \right\}\,,
\end{equation}
with ${\cal A}(\vec{x}) \in [0,1]$. To identify where modes localize, we summed
their amplitude square over time slices, and over modes in the low
($0\le \lambda_n\le \lambda_L$) and high ($\lambda_n\ge \lambda_H$)
spectral regions,
\begin{equation}
  \label{eq:moddens}
  \begin{aligned}
    p_L(\vec{x}) &= \left\la \sum_{\substack{n\\ 0\le \lambda_n\le
          \lambda_L}}\sum_{t=0}^{N_t-1}
      |\psi_n(\vec{x},t)|^2\right\ra\,,
    \\
    p_H(\vec{x}) &= \left\la \sum_{\substack{n\\ \lambda_n\ge
          \lambda_H}}\sum_{t=0}^{N_t-1}
      |\psi_n(\vec{x},t)|^2\right\ra\,,
  \end{aligned}
\end{equation}
with $\lambda_{L,H}$ depending on the center sector under
study. Results are shown in Fig.~\ref{fig:newsi_conf} for one typical
configuration in the real sector and its center-rotated version in the
complex sector $\Im \bar{P}>0$. The correlation between larger
${\cal A}$ and localization is clear in both center sectors. In the
real sector, regions with ${\cal A}$ deviating from zero cover the
areas favorable for localization much more accurately than
fluctuations of the Polyakov loop to the complex sector. In
particular, regions where modes do indeed localize but far from
Polyakov-loop fluctuations are correctly identified by using
${\cal A}\not\approx 0$ as a criterion. In the complex sector, where
the standard sea-islands picture leads one to expect delocalized low
modes, ${\cal A}\not\approx 0$ again correctly identifies locations
where both low and high modes localize. These include also sites where
the Polyakov loop fluctuates to the real sector, which the standard
sea-islands picture would deem ``energetically'' unfavorable. Notice
that on these configurations ${\cal A}$ reaches up at most to around
$0.42$ in the real sector and to around $0.47$ in the complex sector.

To see how modes in different spectral regions respond to fluctuations
in the spatial hopping terms, as measured by ${\cal A}(\vec{x})$, we
have measured the mode weight on sites where this is above a fixed
tolerance level, 
\begin{equation}
  \label{eq:moddensA}
  p_{n}^{({\cal A}_0)} = \sum_{\vec{x}} \sum_{t=0}^{N_t-1} \theta(
  {\cal A}(\vec{x})- {\cal A}_0)|\psi_n(\vec{x},t)|^2\,,   
\end{equation}
and averaged it locally in the spectrum according to
Eq.~\eqref{eq:loc3}. We analyzed separately configurations in the real
and complex Polyakov-loop sectors, also in the confined phase where
the difference should be milder. Results are shown in
Fig.~\ref{fig:newsi_ave}, using ${\cal A}_0=1/16$. Since the volume
dependence is rather mild, only data for $N_s=32$ are shown. We used
200 configurations for each $\beta$ value. Low and high modes favor
regions with larger ${\cal A}(\vec{x})$ in all phases and center
sectors. In the real sector of the deconfined phase,
$\overline{p^{({\cal A}_0)}}(\lambda)$ changes abruptly when entering
the bulk, where it drops by a factor of 2 or more.  In the complex
sector the decrease is smoother, but a change in behavior is clearly
visible, and a significant drop is found comparing the lowest modes
with the bulk modes. Since in the deconfined phase sites with
${\cal A}(\vec{x})\not\approx 0$ become less frequent, low modes
favoring them comes at the price of becoming localized.

These results strongly support the refined sea-islands picture
discussed in Sec.~\ref{sec:newsi}, which allows one to explain the
observed localization of low modes in the complex Polyakov-loop
sectors. This also partially explains the strong correlation between
localized modes and nontrivial plaquettes displayed in
Fig.~\ref{fig:negplaq}.  Indeed, nontrivial spatial-temporal
plaquettes indicate the presence of the kind of disorder in the
hopping terms that, as argued above, leads to favorable locations for
localized low (as well as high) modes.

\section{Conclusions}
\label{sec:concl}

Localized low Dirac modes are found in the deconfined phase of many
gauge theories and related
models~\cite{Giordano:2021qav,Gockeler:2001hr,Gattringer:2001ia,
  Gavai:2008xe,Kovacs:2009zj,Kovacs:2010wx,Bruckmann:2011cc,
  Kovacs:2017uiz,Giordano:2019pvc,Vig:2020pgq,Bonati:2020lal,
  Baranka:2021san,Giordano:2016nuu,Cardinali:2021fpu,
  GarciaGarcia:2006gr,Kovacs:2012zq,Dick:2015twa,Cossu:2016scb,
  Holicki:2018sms,Giordano:2015vla,Giordano:2016cjs,
  Giordano:2016vhx,Bruckmann:2017ywh}, appearing precisely at
deconfinement when this is a genuine phase
transition~\cite{Gockeler:2001hr,Gattringer:2001ia,Gavai:2008xe,
  Kovacs:2009zj,Kovacs:2010wx,Bruckmann:2011cc,Kovacs:2017uiz,
  Giordano:2019pvc,Vig:2020pgq,Bonati:2020lal,Baranka:2021san,
  Giordano:2016nuu,Cardinali:2021fpu}. This naturally suggests a close
connection between low-mode localization and deconfinement. An
explanation of this connection is provided by the sea-islands picture
of localization~\cite{Bruckmann:2011cc,Giordano:2015vla,
  Giordano:2016cjs,Giordano:2016vhx,Giordano:2021qav}, according to
which islands of fluctuations in the sea of ordered Polyakov loops
found in the deconfined phase provide favorable locations for Dirac
eigenmodes, as they effectively and locally reduce the twist on the
fermion wave functions induced by the antiperiodic temporal boundary
conditions. A prediction of the sea-islands picture is then that the
low-lying Dirac modes become localized in the deconfined phase of a
gauge theory, under quite general conditions. So far, this prediction
has always been successfully verified. Moreover, numerical support for
the proposed mechanism has been provided~\cite{Bruckmann:2011cc,
  Cossu:2016scb,Holicki:2018sms,Baranka:2021san}.

In this paper we have studied the localization properties of the
eigenmodes of the staggered lattice Dirac operator in 2+1 dimensional
$\mathbb{Z}_3$ pure gauge theory. This model provides nontrivial tests
for the standard sea-islands picture of localization outlined
above. In the deconfined phase in the physical, real Polyakov-loop
sector where the Polyakov loop gets ordered near $P(\vec{x})=1$,
fluctuations to the complex sectors $P(\vec{x})=e^{\pm i\f{2\pi}{3}}$
provide only a mild gain in twist, and low modes may not be able to
localize. More importantly, in the complex Polyakov-loop sectors where
$P(\vec{x})$ gets ordered near $e^{i\f{2\pi}{3}}$ or
$e^{- i\f{2\pi}{3}}$, local fluctuations provide no gain in twist at
all, leaving it unchanged (for fluctuations to the opposite complex
sector) or even increasing it (for fluctuations to the real sector). A
simple-minded use of the sea-island picture then leads one to expect
that low modes do not localize in this case.

Our numerical results show that localized low modes are present in the
deconfined phase both in the real and in the complex sectors,
appearing right at the deconfinement transition in both cases. While
for the real sector our results agree with the general expectations of
the standard sea-islands picture, for the complex sector this is
unexpected. Even more puzzlingly, in this case the localized low modes
do not avoid Polyakov-loop fluctuations to the real sector, as one
would expect.

A comprehensive understanding of these results is obtained by
reconsidering the sea-islands picture from the point of view of the
spatial hopping terms of the Dirac operator, rather than of the
Polyakov-loop fluctuations. Hopping terms are strongly influenced by
the presence of Polyakov-loop fluctuations, but quite independently of
the gain or loss in the temporal twist on the wave functions that
these provide. Moreover, for a strongly ordered configuration of
Polyakov loops \textit{and} spatial links, the resulting
``ordered-type'' hopping terms lead to the opening of a gap in the
spectrum and to full delocalization of the eigenmodes; deviations from
order modify the hopping terms to ``non-ordered-type'', and generally
allow for eigenvalues below the gap. Typical configurations in the
deconfined phase display a sea of sites connected by ordered-type
hopping terms, with rare islands where one or more of the hopping
terms is of non-ordered-type. Localizing on these islands allows the
mode to penetrate the spectral gap, and so explains localization of
the low modes, as well as their low density. In the language of
first-order perturbation theory, modes living on islands are stable
against delocalization due to the fact that they can hardly mix among
themselves, due to large spatial separation, and with delocalized
modes living on the sea, due to the large energy difference coming
from the presence of a gap.

Islands where hopping terms of non-ordered type are present can also
support modes with much larger eigenvalues than those found in the
presence of ordered-type hopping terms, and so support localized high
modes by a similar no-mixing argument. Even in the confined phase,
where no sea is present, particularly favorable fluctuations in the
hopping terms supporting very large modes are likely to be spatially
separated, and so high modes are again expected to be localized.
However, it is only in the deconfined phase where a spectral
(pseudo)gap opens that low modes living on islands are stable against
delocalization. In the confined phase there is instead no gap and
there are no islands, and so no no-mixing argument and no reason for
low modes not to delocalize. A simple way to describe the different
situations in the two phases is that the ordering of the Polyakov loop
and the resulting spectral pseudogap in the deconfined phase makes the
near-zero region similar to a spectrum edge with low spectral
density. In such a region even a relatively weak disorder (and in
gauge theories the disorder strength is bounded due to the unitary
nature of link variables) can lead to mode localization, as it is
known from the study of Anderson models.

While non-ordered-type islands are expected to correlate strongly with
Polyakov-loop fluctuations away from its ordered value, they do not
require any gain in temporal twist to become favorable to localization
(and can also be found away from any Polyakov-loop fluctuation). This
is consistent with the observed correlation between localized modes
and Polyakov-loop fluctuations in the physical center sector of the
deconfined phase~\cite{Bruckmann:2011cc,Cossu:2016scb,Holicki:2018sms,
  Baranka:2021san}. At the same time, this also explains why low modes
can localize even in the complex sectors of $\mathbb{Z}_3$ gauge
theory in the deconfined phase, where no gain in temporal twist can be
obtained anywhere.

It is worth noticing that in SU(3) pure gauge theory no localized
modes were found in the complex $\mathbb{Z}_3$ center sectors at the
critical point~\cite{Kovacs:2021fwq}, while the results of
Refs.~\cite{Gockeler:2001hr,Gattringer:2001ia} deeper in the
deconfined phase do not allow for conclusive statements. This calls
for further investigation of the onset of low-mode localization in a
complex center sector of a gauge theory.

While the specific results obtained for $\mathbb{Z}_3$ are likely to
be strongly affected by the discreteness of the gauge group and the
lower dimensionality of the system, the refined sea-islands mechanism
unveiled here should be of universal value and apply to a general
gauge theory. This should be tested on physically more relevant
models, including lattice QCD.

\begin{acknowledgments}
  We thank M.~Caselle and A.~Papa for correspondence on the
  $\mathbb{Z}_N$ models, and T.~G.~Kov\'acs for discussions and for a
  careful reading of the manuscript. MG was partially supported by the
  NKFIH grant KKP-126769.
\end{acknowledgments}

\appendix

\section{Duality in 2+1 dimensional $\mathbb{Z}_N$ gauge theories on
  finite lattices}
\label{sec:app_dual}

The partition function of 2+1 dimensional $\mathbb{Z}_N$ gauge
theories on a finite $N_1\times N_2\times N_3$ cubic lattice $\Lambda$
can be written as (see Ref.~\cite{Wipf:2013vp})
\begin{equation}
  \label{eq:dual_simp1}
  Z = e^{-3\beta {\cal V}} \sum_{\{k_p\}} \prod_\ell 
  \delta_{C_\ell,0}
  \prod_p c_{k_p}(\beta)\,,
\end{equation}
where ${\cal V}=N_1 N_2 N_3$, and $\ell$ and $p$ run, respectively, over
links and plaquettes, with links conventionally oriented in the same
direction as the unit lattice vectors, and plaquettes oriented
counterclockwise. The sum over $\{k_p\}$ runs over all choices of the
integers $k_p=0,\ldots,N-1$, each associated with a plaquette
$p$. Moreover, $c_{k_p}$ are known coefficients and, for each link
$\ell$, $\delta_{C_\ell,0} $ imposes the constraint
\begin{equation}
  \label{eq:dual_simp2}
C_\ell=  \sum_{\substack{p\\ \ell\in \de p}} \tau_p k_p =  0\mod N\,,
\end{equation}
where $\de p$ is the boundary of plaquette $p$, and $\tau_p = +1$ or
$-1$ depending on whether one traverses $\ell$ along or opposite to
its orientation when going around $p$.

The constraints in Eq.~\eqref{eq:dual_simp2} are most easily solved
using the dual lattice $\tilde{\Lambda}$, with dual sites located at
the center of elementary cubes of the original (direct) lattice. Dual
links $\tilde{\ell}$ connecting dual sites pierce exactly one of the
direct plaquettes $p$ perpendicularly, and in the same direction as
the plaquette orientation. Dual links and direct plaquettes are then
identified.  In this setup, after setting
$\tilde{k}_{\tilde{\ell}}=k_p$, solving Eq.~\eqref{eq:dual_simp2} is
equivalent to finding the most general configuration of 
gauge link variables
$V_{\tilde{\ell}}=e^{i\f{2\pi \tilde{k}_{\tilde{\ell}}}{N}}$ such that
for all elementary dual plaquettes one has
$\prod_{\tilde{\ell}\in \de \tilde{p}}V_{\tilde{\ell}}=1$. The
solution is found by transforming to the maximal temporal gauge (mtg),
\begin{equation}
  \label{eq:dual_simp3}
  \begin{aligned}
    V^{\rm mtg}_{\tilde{\ell}}&=1\text{ for }\\
    \tilde{\ell}&= \left\{
      \begin{aligned}
        &(\tilde{n},\hat{1}), &&& &0\le \tilde{n}_1 < N_1-1\,;\\
        &(\tilde{n},\hat{2}), ~\tilde{n}_1=0, &&& &0\le \tilde{n}_2 < N_2-1\,;\\
        &(\tilde{n},\hat{3}), ~\tilde{n}_{1,2}=0, &&& &0\le \tilde{n}_3 < N_3-1\,,
      \end{aligned}
    \right.
  \end{aligned}
\end{equation}
where $\tilde{\ell}=(\tilde{n},\hat{\mu})$ is the dual link connecting
$\tilde{n}$ and $\tilde{n}+\hat{\mu}$. For each configuration there
are exactly $N$ gauge transformations $G_g(\tilde{n})$, all leading to
the same set of new link variables $V^{\rm mtg}_{\tilde{\ell}}$
satisfying the maximal temporal gauge condition
Eq.~\eqref{eq:dual_simp3},
\begin{equation}
  \label{eq:dual_simp4}
  V_{\tilde{\ell}} =   G_g(\tilde{n})\, V^{\rm
    mtg}_{\tilde{\ell}} \,G_g(\tilde{n}+\hat{\mu})^*\,.
\end{equation}
These read $G_g(\tilde{n})= e^{i\f{2\pi g}{N}}s(\tilde{n})$, with
$g = 0,\ldots, N-1,$ and
$s(\tilde{n})=e^{i\f{2\pi \sigma(\tilde{n})}{N}}$, with
$\sigma(\tilde{n})=0,\ldots,N-1$, and
\begin{equation}
  \label{eq:tgauge_sol}
  \begin{aligned}
    s(\tilde{n})&= W_3(0,0,0;0,0,\tilde{n}_3)\\ &\phantom{=}\times
    W_2(0,0,\tilde{n}_3;0,\tilde{n}_2,\tilde{n}_3)\\
    &\phantom{=}\times
    W_1(0,\tilde{n}_2,\tilde{n}_3;\tilde{n}_1,\tilde{n}_2,\tilde{n}_3)
    \,,\\
    W_\mu(\tilde{n};\tilde{n}+L\hat{\mu}) &=
    \prod_{s=0}^{L-1}V_{(\tilde{n}+s\hat{\mu},\hat{\mu})}{}^*\,.
  \end{aligned}
\end{equation}
In this gauge the solution is readily found and reads
\begin{equation}
  \label{eq:dual_simp5}
  \begin{aligned}
    V_{\tilde{\ell}}^{\rm mtg}&=1\,, &&& \text{if }&\tilde{\ell}\not\in
    \cup_{\mu=0}^3\de_\mu\tilde{\Lambda}\,,\\
    V_{\tilde{\ell}}^{\rm mtg}&=B_\mu=e^{i\f{2\pi b_\mu}{N}}\,, &&& \text{if
    }&\tilde{\ell}\in \de_\mu\tilde{\Lambda}\,,
  \end{aligned}
\end{equation}
with $b_\mu=0,\ldots,N-1$, and where
\begin{equation}
  \label{eq:dual_simp6}
  \de_\mu\tilde{\Lambda}=\{\tilde{\ell}=(\tilde{n},\hat{\mu})~|~\tilde{n}_\mu
  = N_\mu-1\} 
\end{equation}
denotes the links on the boundary of the dual lattice in direction
$\mu$. The value of $B_\mu$ is the same across the whole boundary
$\de_\mu\tilde{\Lambda}$. Undoing the gauge transformation, one writes
for the most general solution
\begin{equation}
  \label{eq:dual_simp_solution}
 V_{(\tilde{n},\hat{\mu})}=e^{i\f{2\pi
    \tilde{k}_{(\tilde{n},\hat{\mu})}}{N}}=s(\tilde{n})s(\tilde{n}+\hat{\mu})^*\,, 
\end{equation}
with arbitrary $s(\tilde{n})$ obeying the boundary condition
$s(\tilde{n}+N_\mu\hat{\mu})=B_\mu s(\tilde{n})$. Equivalently, one
has
$\tilde{k}_{(\tilde{n},\hat{\mu})} =
\sigma(\tilde{n})-\sigma(\tilde{n}+\hat{\mu})$, with arbitrary
$\sigma(\tilde{n})$ satisfying the boundary condition
$\sigma(\tilde{n}+N_\mu\hat{\mu})=\sigma(\tilde{n})+b_\mu \mod N$. One
easily shows that the $V_{(\tilde{n},\hat{\mu})}$ are uniquely
identified by the spin variables $s(\tilde{n})$ or $\sigma(\tilde{n})$
and by the boundary conditions $B_\mu$ up to global transformations
$s(\tilde{n})\to s(\tilde{n})e^{i\f{2\pi g}{N}}$, $g=0,\ldots,N-1$,
corresponding to the $N$ gauge transformations leading to maximal
temporal gauge. Summing without restrictions over $s(\tilde{n})$ or
$\sigma(\tilde{n})$ and over all possible boundary conditions yields
then all the allowed configurations of $\tilde{k}_{\tilde{\ell}}$,
with each configuration appearing exactly $N$ times. One concludes
that
\begin{equation}
  \label{eq:dual_simp_final}
  \begin{aligned}
    Z &
    =  e^{-3\beta {\cal V}} N^{-1}\sum_{\{b_\mu\}} \tilde{Z}_{\{b_\mu\}}\,,\\
    \tilde{Z}_{\{b_\mu\}}&= \sum_{\{\sigma(\tilde{n})\}}
    \prod_{(\tilde{n},\hat{\mu})} c_{k_{(\tilde{n},\hat{\mu})}}(\beta)
    \Big|_{\substack{k_{(\tilde{n},\hat{\mu})}=\sigma(\tilde{n})-\sigma(\tilde{n}+\hat{\mu})\,{\rm
          mod}\,
        N\\\sigma(\tilde{n}+N_\mu\hat{\mu})=\sigma(\tilde{n})+b_\mu
        \,{\rm mod}\, N}} \,,
  \end{aligned}
\end{equation}
which is the desired duality relation. Substituting the values of
$c_{k}(\beta)$ for $N=3$ one finds Eq.~\eqref{eq:z3_3}.

\section{Sea-islands picture: technical details}
\label{sec:seaislands}

\subsection {Ordering of on-site energies}
\label{sec:seaislands_rank}

The ranking of the unperturbed energy levels
$e_k(\vec{x})= \eta_{d+1}(\vec{x})\sin\omega_k(\vec{x})=
\eta_{d+1}(\vec{x})\sin\tilde{\omega}_{N_k(\vec{x})}(\phi(\vec{x}))$
by magnitude [see Eqs.~\eqref{eq:DA_def5} and \eqref{eq:DA7_alt}] is
achieved by setting
$N_k(\vec{x})=n_k(\phi(\vec{x}),\eta_{d+1}(\vec{x}))$, with
$n_k(\phi,\eta_{d+1})$ chosen as follows:
\begin{equation}
  \label{eq:si0}
  \begin{aligned}
    n_{2k}(\phi,1)&= \theta_0(-\phi) k +
    \theta_0(\phi)\left(\tf{N_t}{2}-1- k\right)\,,
    \\
    n_{2l+1}(\phi,1)&=\theta_0(-\phi)\left(\tf{N_t}{2}-1- l\right)
    +\theta_0(\phi)l\,,
    \\
    n_{2k}(\phi,-1)&= \tf{N_t}{2} +n_{2k}(\phi,1)\,,
    \\
    n_{2l+1}(\phi,-1)&= \tf{N_t}{2}+ n_{2l+1}(\phi,1)\,,
  \end{aligned}
\end{equation}
with $k,l\in \{0,\ldots,\tf{N_t}{4}-1\}$ if $N_t=4m$, and
$k\in \{0,\ldots,\tf{N_t-2}{4}\}$, $l\in \{0,\ldots,\tf{N_t-2}{4}-1\}$
if $N_t=4m+2$, and
\begin{equation}
  \label{eq:si1}
  n_{\f{N_t}{2}+ k}(\phi,\pm 1)=  \tf{N_t}{2}+  n_{k}(\phi,\pm
  1) \mod N_t\,,
\end{equation}
with $k\in \{0,\ldots,\tf{N_t}{2}-1\}$. Here $\theta_0(x)=1$ if
$x\ge 0$ and $\theta_0(x)=0$ if $x<0$.

\subsection{Strongly ordered configurations}
\label{sec:seaislands_soc}

For strongly ordered configurations with spatially
constant Polyakov loop,
$P(\vec{x})= P_*=e^{i\phi_*}$, the quantity
$\omega_k(\vec{x})=\tilde{\omega}_{n_k(\phi_*,\eta_{d+1}(\vec{x}))}(\phi_*)\equiv
\tilde{\omega}_{\hat{n}_k(\eta_{d+1}(\vec{x}))}$ depends on $\vec{x}$
only through the $\eta_{d+1}$ dependence of $n_k$, and so only on
whether $\vec{x}$ is an even or odd site ($\eta_{d+1}=\pm1$).
Moreover, from Eqs.~\eqref{eq:si0} and \eqref{eq:si1} one has
\begin{equation}
  \label{eq:si2}
  \hat{n}_k(\pm 1)=\f{N_t}{2}+\hat{n}_k(\mp 1) \mod N_t\,.
\end{equation}
If also
$U_{\pm j}^{\rm tg}(\vec{x},t)= U_{\pm j *}^{\rm tg}(\vec{x})$, as one
would approximately expect when there are strong temporal correlations
and spatial-temporal plaquettes ($\mu=j$, $\nu=d+1$) reduce to
$U_{j\, d+1}(\vec{x},t) \approx U_j(\vec{x},t)U_j(\vec{x},t+1)^*$,
then from Eqs.~\eqref{eq:DA_def4}, \eqref{eq:DA_def5} and
\eqref{eq:si2} one finds
\begin{equation}
  \label{eq:si3}
  \begin{aligned}
    V_{\pm j}(\vec{x})_{kl} &= U_{\pm j *}^{\rm tg}(\vec{x}) \\
    &\phantom{=}\times \f{1}{N_t}\sum_{t=0}^{N_t-1}e^{-i\f{
        2\pi}{N_t}\left(\hat{n}_k(\eta_{d+1}(\vec{x}))
        -\hat{n}_l(\eta_{d+1}(\vec{x}))- \f{N_t}{2} \right)t} \\ &=
    U_{\pm j *}^{\rm tg}(\vec{x}) \delta_{k,l+\f{N_t}{2}\,{\rm mod}\,
      N_t}\,.
  \end{aligned}
\end{equation}
If a perfect anticorrelation was found for the spatial links,
$U_{\pm j}^{\rm tg}(\vec{x},t)= (-1)^tU_{\pm j *}^{\rm tg}(\vec{x})=
e^{i\pi t}U_{\pm j *}^{\rm tg}(\vec{x})$, then
\begin{equation}
  \label{eq:si3bis}
  \begin{aligned}
    V_{\pm j}(\vec{x})_{kl} &= U_{\pm j *}^{\rm tg}(\vec{x}) \\
    &\phantom{=}\times \f{1}{N_t}\sum_{t=0}^{N_t-1}e^{-i\f{
        2\pi}{N_t}\left(\hat{n}_k(\eta_{d+1}(\vec{x}))
        -\hat{n}_l(\eta_{d+1}(\vec{x})) \right)t} \\ &= U_{\pm j
      *}^{\rm tg}(\vec{x}) \delta_{k,l}\,.
  \end{aligned}
\end{equation}
These results differs from those reported in
Ref.~\cite{Giordano:2016cjs} due to the different convention used in
defining $\omega_k(\vec{x})$, in particular the inclusion of
$\eta_{d+1}$ in the quantities to be ranked.

\subsection{Non-Abelian case}
\label{sec:si_NA}

Here we extend the argument of Sec.~\ref{sec:newsi} to a non-Abelian
theory, with link variables $U_\mu(n)$ providing a unitary
representation of the gauge group (assumed to be semisimple and
compact). In this case the unperturbed eigenvalues $e_{ka}(\vec{x})$
have a further index $a=1,\ldots, N_c$, running over the internal
``color'' degree of freedom, and are obtained as
\begin{equation}
  \label{eq:DA4_alt_NA}
  \begin{aligned}
    e_{ka}(\vec{x})&=    \eta_{d+1}(\vec{x})\sin\omega_{ka}(\vec{x})\,,\\
    \omega_{ka}(\vec{x}) &=
    \tilde{\omega}_{N_{ka}(\vec{x})}(\phi_a(\vec{x}))\,,
  \end{aligned}
\end{equation}
with $\phi_a(\vec{x})\in[-\pi,\pi)$ the $N_c$ eigenphases of the
Polyakov loop
\begin{equation}
  \label{eq:DA4_alta_NA2}
  \begin{aligned}
    P(\vec{x}) &= u(\vec{x})^\dag {\rm
      diag}(e^{i\phi_1(\vec{x})},\ldots,e^{i\phi_{N_c}(\vec{x})})u(\vec{x})\,,\\
    u(\vec{x})^\dag u(\vec{x})&=\mathbf{1}\,.
  \end{aligned}
\end{equation}
Notice that with our convention one generally finds for special
unitary $U_\mu(n)$ that $\sum_{a=1}^{N_c}\phi_a = 2\pi q$ with integer
but not necessarily zero $q$. This differs from the choice made in
Ref.~\cite{Giordano:2016cjs}. The hopping matrices also acquire extra
indices, $V_{\pm j}(\vec{x})_{ka\,lb}$, and are now defined as
\begin{equation}
  \label{eq:DA_def4_NA}
  \begin{aligned}
    V_{\pm j}(\vec{x})_{ka\, lb}&=
    \f{1}{N_t}\sum_{t=0}^{N_t-1}e^{-i[\omega_{ka}(\vec{x}) -
      \omega_{lb}(\vec{x}\pm\hat{\jmath})]t} [U_{\pm j}^{\rm
      tdg}(\vec{x},t)]_{ab}\,,\\
    U_{\pm j}^{\rm tdg}(\vec{x},t) &=P(\vec{x},t) U_{\pm j}(\vec{x},t)
    P(\vec{x}\pm \hat{\jmath},t)^\dag\,,
  \end{aligned}
\end{equation}
where $P(\vec{x},t+1)= P(\vec{x},t) U_{d+1}(\vec{x},t)$,
$P(\vec{x},0)=\mathbf{1}$ is the $N_c$-dimensional identity matrix,
$P(\vec{x},N_t) =P(\vec{x})$, and moreover
$U_{-j}(\vec{x},t) = U_j(\vec{x}-\hat{\jmath},t)^\dag$. Here ``tdg''
denotes the temporal diagonal gauge where
$U_{d+1}^{\rm tdg}(\vec{x},t)=\mathbf{1}$, $\forall\vec{x}$,
$0\le t< N_t-1$, and all Polyakov loops are diagonal,
$ P^{\rm tdg}(\vec{x})= {\rm diag}(e^{i\phi_a(\vec{x})})$.  Notice
that $V_{\pm j}(\vec{x})$ are now unitary matrices in the extended
$N_cN_t$-dimensional space. As long as $N_{ka}(\vec{x})$ is chosen so
that Eq.~\eqref{eq:oppo} holds for all $a$, i.e.,
$e_{k+\f{N_t}{2}{\rm mod}\,N_t \,a}(\vec{x})=-e_{ka}(\vec{x})$, and
that $e_{ka}(\vec{x})\ge 0$ for $k=0,\ldots,\f{N_t}{2}-1$, then
Eq.~\eqref{eq:DA_general} holds, and the argument outlined in Section
\ref{sec:newsi} carries through.

\bibliographystyle{apsrev4-2}
\bibliography{references_gt}

\end{document}